# Inferring directed spectral information flow between mixed-frequency time series

Qiqi Xian[a,b] and Zhe Sage Chen[a,c,d,*]

a. Department of Psychiatry, Department of Neuroscience and Physiology, New York University Grossman School of Medicine, New York, NY 10016, USA

b. School of the Gifted Young, University of Science and Technology of China, Hefei, Anhui 230022, China

c. Neuroscience Institute, New York University Grossman School of Medicine, New York, NY 10016, USA

d. Department of Biomedical Engineering, New York University Tandon School of Engineering, Brooklyn, NY 11201, USA

[*] To who correspondence may be addressed. Email: zhe.chen@nyulangone.org (Z.S.C.)

ORCID:
0009-0004-6859-0091 (Q. Xian)
0000-0002-6483-6056 (Z.S. Chen)

**Author contributions:** Z.S.C. conceived and supervised experiments, developed the methods, interpreted the data, and wrote the paper. Q.X. developed the methods, performed experiments, analyzed and interpreted the data. Z.S.C. acquired funding.

**Competing interest statement:** The authors declare no competing interest.

Number of Figures: 8
Number of Box: 1
Number of Supplementary Figures: 10
Number of Supplementary Tables: 2

**Abstract**

**Identifying directed spectral information flow between multivariate time series is important for many applications in finance, climate, geophysics and neuroscience. Spectral Granger causality (SGC) is a prediction-based measure characterizing directed information flow at specific oscillatory frequencies. However, traditional vector autoregressive (VAR) approaches are insufficient to assess SGC when time series have mixed frequencies (MF) or are coupled by nonlinearity. Here we propose a time-frequency canonical correlation analysis approach ("MF-TFCCA") to assess the strength and driving frequency of spectral information flow. We validate the approach with extensive computer simulations on MF time series under various interaction conditions and further assess statistical significance of the estimate with surrogate data. In various benchmark comparisons, MF-TFCCA consistently outperforms the traditional parametric MF-VAR model in both computational efficiency and detection accuracy, and recovers the dominant driving frequencies. We further apply MF-TFCCA to real-life finance, climate and neuroscience data. Our analysis framework provides an exploratory and computationally efficient nonparametric approach to quantify directed information flow between MF time series in the presence of complex and nonlinear interactions.**

## INTRODUCTION

Quantifying directed information flow between two or more simultaneously measured time series is a common problem in science and engineering, including economics, finance, meteorology, transportation, geophysics, and neuroscience[1-4] (Granger, 1969; Surgihara et al., 2012; Silva et al., 2021; Yuan and Shou, 2022). These time series are sometimes collected at different temporal resolution. For instance, this may occur in finance by comparing a daily stock price with a weekly trade transaction; or may occur in weather forecast by linking monthly rainfall amount at the city level to yearly rainfall at the national level; or may occur in neuroscience when concurrent neural recordings of different modalities (e.g., calcium imaging and electrophysiology; concurrent EEG-fMRI) are simultaneously collected (**Fig. 1a**).

Statistical dependency of two time series may be characterized in the time domain (such as lagged cross-correlation, or "xcorr") or in the frequency domain (such as coherence), which bears many other names in different contexts, such as functional connectivity[5] (Bastos and Schoffelen, 2016), partial directed coherence (PDC)[6] (Baccala and Sameshima, 2001), Granger causality (GC) or GC-like measures[7-11] (Granger, 1980; Geweke, 1984; Ding et al., 2006; Barnett and Seth, 2014; Shojaie and Fox, 2022), and transfer entropy[12] (Schreiber, 2000). These measures are either undirected or directed, where the direction implies an asymmetry in information flow between the "source" and "sin" (or "target"). Two generalized scenarios are noteworthy. First, multiple (greater than two) time series are observed; in this case, multivariate statistical methods have been developed to assess their conditional dependence (e.g., conditional GC, partial correlation, partial directed coherence). Second, two time series have distinct temporal resolution; in this case, different analysis strategies are required[11] (Shojaie and Fox, 2022). One naive strategy is to apply subsampling or temporal averaging to the signal with higher sampling frequency to match the one with lower sampling frequency, and then to characterize their interactions based on standard metrics. Another strategy originated from the econometrics literature is GC testing with mixed frequency (MF) data using a MF-VAR model[13-15] (Ghysels et al., 2016; Gotz and Hecq, 2019; Anderson et al., 2016). The MF-VAR method, however, only works well when the discrepancy of MF is small or the lag is relatively small, and cannot account for nonlinear interactions. Several lines of work have recently studied the identifiability of a structural VAR(1) model under arbitrary subsampling[16] (Gong et al., 2015) and MF[17] (Tank et al., 2019) settings. The nonlinear interactions have also been studied in the framework of non-Gaussian structural VAR model (e.g., REF[18] Hyvarinen et al., 2010). To our best knowledge, however, these approaches rely on parametric model estimation (such as REF[16,17] Gong et al., 2015; Tank et al., 2019), and cannot recover the driving frequency in the SGC method.

In this paper, we propose a computationally efficient, nonparametric approach to address the above-mentioned challenges, identify the prominent driving frequency (in SGC sense), and assess directed spectral information flow between MF time series. Our approach is motivated from canonical correlation analysis (CCA) in the time-frequency (TF) domain but adapted for MF signals, which we coin "MF-TFCCA" to distinguish from the traditional canonical GC[19,20] (Sato et al., 2010; Ashrafulla et al., 2013). To validate our approach, we conduct systematic computer simulations on bivariate or trivariate VAR processes (with known SGC ground truth) to validate our analysis pipeline, and further apply MF-TFCCA to three real-life datasets. In the case of neuroscience data, our analysis results provide insights into the directed neural interactions between the mouse hippocampus and downstream cortical structures, and further reveal directed hippocampal-neocortical information flow before, during and after hippocampal sharp wave ripples (SWRs) in sleep.

**A methodological overview of GC and directed information flow**

Correlation is the most statistical measure that characterizes linear dependence of two random variables. In the frequency domain, the complex-valued coherency corresponds to the Fourier transform of cross-correlation, and the modulus of the coherency is coherence. While correlation and causation can coexist, correlation does not imply causation. Additionally, two variables can be statistically uncorrelated even if one is caused by the other based on nonlinear transformation (e.g., $y = x^2$). Therefore, the relationship between causality and correlation depends on multiple factors, such as linearity and Gaussianity.

Identifying causality in a complex physical or physiological system is an important task. In such systems, the recorded signals may have single or multiple dominant oscillatory components in the frequency domain. The notion of GC is different from causality in the physical sense. GC or GC-like methods provide a framework that uses temporal predictability as opposed to correlation to identify causation between time-series variables. Specifically, a variable $X$ is said to "Granger cause" a variable $Y$ (notation: as $X \rightarrow Y$) if the predictability of the target variable $Y$ declines when the source variable $X$ is removed from the universe of all possible causative variables[9] (Ding et al., 2006). GC can include both lagged and instantaneous effects between time series. This assumes that causes can be separated from effects, so that a variable is identified as causative if predictability declines when that variable is removed. Here, we restrict the GC effect on the history information only and exclude the instantaneous causal effect. Although GC in the original definition is in a general sense[7] (Granger, 1980), its practical use is often based on linear predictability assumption and Gaussian assumption, and has been facilitated by well-documented software[10] (Barnett and Seth, 2014). Within the VAR framework, directed GC $X \rightarrow Y$ in the time domain can be also expressed equivalently in terms of SGC in the frequency domain, which displays as a high amplitude at a single driving frequency (or the most prominent driving frequency among several modes). The driving frequency is usually linked to the inherent oscillatory frequencies of the "source" signal. In a general setting of bidirectional GC (denoted as $X \rightleftharpoons Y$), two driving frequencies in $X \rightarrow Y$ and $Y \rightarrow X$ directions are different, and $X$ and $Y$ can play both the roles of source and target but at different timescales. GC is a special case of transfer entropy under the linear Gaussian assumption, which defines a direct information transfer between jointly dependent random processes[12,21,22] (Schreiber, 2000; Barnett et al., 2009; Stramaglia et al., 2021).

Since GC is a prediction-based measure, it is generally impossible to determine GC from the asymmetric shape of "xcorr" between two random variables in the time domain[23] (Borysov and Balatsky, 2014). Lagged cross-correlation is neither sufficient nor necessary condition for GC (see ***Supplementary Fig. 1*** for simple examples). However, the lead-lag relationship may provide a hint to the directionality of information flow (but not the driving frequency). Notably, cross-correlation function is influenced not only by the relationship between the signals but also by the autocorrelation of the signals. In other words, autocorrelation within the target variable can also help prediction, leading to a spurious GC from the regressor to the target variable. Therefore, signal prewhitening may override the limitation of revealing the causal relationship between random variables[24] (El-Gohary and McNames, 2007). CCA has also been used to determine *linear* Granger causal relationship[19,20,25] (Sato et al., 2010; Ashrafulla et al., 2013; Zhuang et al., 2020). A naïve canonical GC approach is to maximize the canonical correlation (CC) between eigen-time series at different time lags while removing the influence of autocorrelation within the target time series[19] (Sato et al., 2010). Notably, there is a conceptual difference between CCA-based GC analysis and VAR-based GC analysis in that the former approach is based on two sets of variables, and the latter approach allows a linear combination of more than one variable in the regressor.

Finally, it is uncommon to observe that time series have different temporal resolution induced by subsampling or due to data acquisition discrepancy. Traditional GC approaches do not work for signals with dramatically different temporal resolution. It is also well known that subsampling may produce a bias for the GC estimate[26-28] (Solo, 2007; Barnett and Seith, 2017; Anderson et al., 2019). Several methods have been developed to account for nonlinear and non-Gaussian observations and allow for subsampled and MF time series[11] (Shojaie and Fox, 2022), but remains unclear how various MF and nonlinear coupling conditions may affect the bidirectional SGC estimate.

## Conceptual ideas of our approach

To gain intuition, let's start our reasoning with a simple example. Let's consider $X$ and $Y$ as two univariate random variables from respective time series $X(t)$ and $Y(t)$, with the known ground truth: $X$ Granger causes $Y$ at a driving oscillatory frequency $f_x$, denoted as $X \overset{f_x}{\rightarrow} Y$. From a linear system perspective, a directed information flow $X \rightarrow Y$ suggests that the history of $X(t)$ improves linear predictability of $Y(t)$, or $Y$ contains linearly filtered information from $X$. However, if the interaction component acts like a high-pass filter, the causal information transfer $X \rightarrow Y$ will be largely lost after down-sampling of $Y$ because the autocovariance structure of $Y$ does not contain all information required to estimate the original VAR model. Therefore, the temporal causal relationship in the high-frequency range may not be recovered.

When there is a unidirectional GC relationship $X \rightarrow Y$ and two variables $X$ and $Y$ have MF, we consider two possibilities: the sampling rate of $X$, denoted by $F_x$, is greater than the sampling rate of $Y$, denoted by $F_y$; or the opposite, namely $Y$ has a higher sampling rate than $X$ such that $F_y > F_x$. In both cases, if $F_x < 2f_x$, then it is impossible to recover the causal relationship in light of Nyquist's sampling theorem. Only if $F_x > 2f_x > F_y$, inferring the driving frequency $f_x$ becomes possible. When there is a bidirectional GC relationship $X \rightleftharpoons Y$ and two directions have distinct driving frequencies, the outcome may become more complex and depend on multiple factors, which will be illustrated in the subsequent Results section.

The problem of identifying directed spectral information flow consists of three subtasks (**Fig. 1b**): 1) inferring directionality of information flow; 2) identifying the driving frequency; and 3) quantifying the relative information flow metric (normalized by self-predictability in analogy to autocorrelation) and assessing it statistical significance (related to surrogate data). For the first subtask, we first use MF-TFCCA to compute the CC between a time series and a time-frequency process, where the time-frequency representation (TFR) can be computed by a moving window short-time Fourier transform (STFT). We further determine the directionality based on the asymmetric shape of lag-CC derived from MF-TFCCA by moving the time lag between two time series. If the peak of lag-CC curve deviates from zero (at either positive or negative lag), it implies that the history of one variable can improve the predictability of the other variable. For the second subtask, the driven frequency (or frequencies) may be recovered by the dominant peak (or peaks) of CCA coefficients at the frequency axis. Furthermore, the contribution of individual driving frequencies can be assessed by the relative CC gain based on the "filter-one-frequency-out" strategy. For the third subtask, we resort on permutation statistics using surrogate data[4,10,29] (Yuan and Shou, 2022; Barnett and Seth, 2014; Lancaster et al., 2018). The purpose of the surrogate data is to destroy the signal's temporal regularity by introducing randomness in phase while maintaining the magnitude in the Fourier domain.

## RESULTS

### Analysis pipeline

We first consider the condition where $X(t)$ and $Y(t)$ are two univariate time series, and then relax the assumption to more than two time series. Without the loss of generality, unless specified otherwise, we assume that $X(t)$ has a higher sampling rate than $Y(t)$, and call $X(t)$ as a high-frequency (HF) process, and $Y(t)$ as a low-frequency (LF) process. The MF-TFCCA analysis pipeline (**Fig. 1c**) is described in Box 1.

***Step 1***: We convert univariate $X(t)$ in the time domain into a multivariate TFR $\widetilde{\mathbf{X}}(t,\omega)$ in the time-frequency domain (such as STFT) to match the sampling rate of $Y(t)$. The window size $W_l$ of STFT is chosen based on the sampling rate of $X$ and desired frequency resolution: the longer the window, the higher frequency resolution. The overlap of moving window is $(W_l - 1/F_y)$. For calibration, we apply z-score normalization on each frequency of the complex-valued time-frequency matrix $\widetilde{\mathbf{X}}(t,\omega)$.

*Note*: the z-score normalization on each frequency implies prewhitening of $\widetilde{\mathbf{X}}$ in the frequency-axis but not in the time-axis. The degree of autocorrelation in the time-axis depends on the size of overlapping window: a larger overlap implies a higher degree of autocorrelation.

***Step 2***: We use regularized CCA to assess CC where the linear projection of $\widetilde{\mathbf{X}}$ is maximally correlated with $Y$, from which we compute the frequency vs. CCA coefficient (absolute value) and identify a canonical frequency (say $f_0$) that is associated with the peak CCA coefficient. The canonical frequency often corresponds to the dominant oscillatory frequency $f_x$ in the source signal $X$. The value of CC depends on both temporal and frequency resolution of $\widetilde{\mathbf{X}}$ (i.e., the autocorrelation of the source signal) as well as the temporal resolution of $Y$. To calibrate the CC value, we may normalize CC with an autocorrelation-like CC factor, which is derived from MF-TFCCA between $\widetilde{\mathbf{X}}$ and a down-sampled version of $X(t)$.

*Note 1:* CC will be computed between real-valued $Y$ and complex-valued $\widetilde{\mathbf{X}}$=STFT($X$). We may consider two options for assessing CC and CCA coefficients. Option 1: first applying complex-valued CCA and then taking the magnitude of complex-valued CCA coefficients (CC is always real-valued). Option 2: first taking the absolute value $\mathrm{abs}(\widetilde{\mathbf{X}}) = \mathrm{abs}(\mathrm{STFT}(X))$ and then applying standard CCA to obtain real-valued CCA coefficients and CC.

*Note 2:* As a sanity check, if we filter out the dominant oscillatory frequency $f_x$ in the source signal $X$ (i.e., “filter-one-frequency-out”) and repeat the same procedure, the resultant CC shall be significantly reduced (i.e., a negative CC gain). Therefore, the spectral contributions of individual frequencies can be assessed by comparing their relative CC gains (compared to the original unfiltered setting).

***Step 3***: We shift the time series by a lag parameter and repeat the above procedure to construct a lag-CC curve. The lag-CC curve is usually asymmetric with respect to 0. A large CC value at negative lag indicates that $X \to Y$ in information flow, whereas a large CC value at positive lag indicates an opposite direction $Y \to X$. If the lag-CC curve is approximately symmetric, an additional step (such as “filter-one-frequency-out”) is required to identify whether causality is bidirectional.

*Step 4*: To assess the statistical significance of lag-CC, we create the surrogate data of both $X$ and $Y$, and generate a null distribution of CC. If the original CC value is significantly greater than the surrogate data statistics, a significance outcome will be reported (t-test's *P*-value for multiple trials, or two-sample Kolmogorov-Smirnov test's *P*-value). By comparing the lag-CC profiles derived from the original and surrogate data, we may determine the directed information flow between $X$ and $Y$.

In the following sections, we will validate MF-TFCCA using computer simulations and real-world datasets. A summary of computer simulation results is given in **Supplementary Table 1**.

**Computer Simulations**

**Bivariate bidirectional systems with distinct driving frequencies**. We started with a bivariate and bidirectional GC system $X \rightleftharpoons Y$, $X \overset{f_x}{\rightarrow} Y$ and $Y \overset{f_y}{\rightarrow} X$. The system has two variables that mutually excite each other at different time lags at two different frequencies: $f_x = 4$ Hz and $f_y = 15$ Hz (**Fig. 2a**). The original time series were generated from a bivariate 4th-order AR process with a sampling rate of 200 Hz (**Methods**), with a total of 100 trials and 20 s per trial (**Fig. 2b**). After down-sampling of $Y$ by a factor of 5, two time series had distinct temporal resolution: $F_x =$ 200 Hz and $F_y =$ 40 Hz. At the first examination of lag cross-correlation (“xcorr”), there was two noticeable positive peaks: one peak at negative lag of ~150 ms and the other peak at positive lag of ~100 ms (**Fig. 2c**). To compute the STFT of $X$, the window length was chosen to be 200 ms, and the overlap for the moving window was chosen as 200-25=175 ms. Consequently, the time-frequency matrix $\tilde{\mathbf{X}}$ had the same length in the time-axis as the time series vector $Y$. In this case, MF-TFCCA recovered similar lag-CC shapes as in xcorr: the peak at the negative lag indicates $X \rightarrow Y$ and the peak at the positive lag indicates $Y \rightarrow X$ (**Fig. 2d**). Furthermore, examining the absolute CCA coefficients showed a peak at 4 Hz in the $X \rightarrow Y$ direction, and a high peak at 15 Hz in the $Y \rightarrow X$ direction. We also witnessed an energy leakage at 4 Hz because of the close distance between two driving frequencies.

Although in this example, MF-TFCCA generated similar GC results consistent with xcorr and recovered the driving oscillatory frequencies. This observation could change depending on the relationship of driving frequency and sampling frequency. To illustrate this point, we also considered a special case when GC is unidirectional (either $X \overset{f_x}{\rightarrow} Y$ with $f_x = 80$ Hz and 4 Hz or $Y \overset{f_y}{\rightarrow} X$ with $f_y = 15$ Hz; **Methods**). We assumed that $X(t)$ had intrinsic oscillations at 4 Hz and 80 Hz, whereas $Y(t)$ had intrinsic oscillations at 2 Hz and 15 Hz (***Supplementary Fig. 2a***). First, we tested the direction of $X \overset{f_x}{\rightarrow} Y$ and down-sampled $Y$ by a factor of 5 (***Supplementary Fig.***

**2b**). We computed the STFT of $X$ using a window size of 150 ms (***Supplementary Fig. 2c***). In this case, the xcorr profile showed an asymmetric shape between the original $X(t)$ and subsampled $Y(t)$ (***Supplementary Fig. 2d***). We also recovered an asymmetric lag-CC profile, suggesting that the direction of information flow is from $X$ to $Y$ (***Supplementary Fig. 2e***). The CCA coefficients showed two peaks at 4 Hz and 80 Hz (***Supplementary Fig. 2f***), which matched two dominant driving frequencies. Note that although the SGC $X \rightarrow Y$ was higher in 80 Hz than in 4 Hz, the causal relationship in the high-frequency band was suppressed after down-sampling of $Y$, resulting in a lower peak of CCA coefficients in 80 Hz than in 4 Hz. A follow-up "filter-one-frequency-out" analysis revealed the *relative* CC contributions of these two driving frequencies: a large negative CC gain suggests an important contribution of the putative driving frequency (i.e., 4 Hz) within the filtered-out frequency band, whereas a nearly zero CC gain suggests negligible contribution of the other putative driving frequency (***Supplementary Fig. 2g***). The reason why we didn't observe a clear negative CC gain at 80 Hz is probably the spectral causality was main preserved in the low frequency range after down-sampling of $Y$.

Further, we tested the direction of $Y \overset{f_y}{\rightarrow} X$ and down-sampled $X$ by a factor of 5 (***Supplementary Fig. 2h***). In this case, xcorr showed a nearly symmetric shape (***Supplementary Fig. 2i***) and MF-TFCCA uncovered the true GC relationship and information flow from $Y$ to $X$. Importantly, comparing the $Y \rightarrow X$ and $X \rightarrow Y$ directions, the estimated lag-CC values (0.8 vs. 0.4; ***Supplementary Fig. 2j vs. Supplementary Fig. 2e***) also quantitatively preserved the relationship of relative SGC values (5 vs. 2.5; ***Supplementary Fig. 2h vs. Supplementary Fig. 2a***). The CCA coefficients recovered one dominant peak at the 15 Hz driving frequency and second peak at 80 Hz (false positive) (***Supplementary Fig. 2k***). Similarly, a subsequent "filter-one-frequency-out" analysis revealed the relative CC contribution between these two putative driving frequencies (***Supplementary Fig. 2l***).

Next, we investigated the impact of down-sampling factor of $Y$ on the lag-CC estimate. Using the above unidirectional system as an example ($X \overset{f_x}{\rightarrow} Y$, $f_x = 80$ Hz and 4 Hz), we systematically varied the sampling frequency of $Y$ while keeping the sampling rate of $X$ intact and used MF-TFCCA based on different window sizes. To accommodate the change in ratio $\frac{F_x}{F_y}$, we adapted the overlap of the moving window to assure the same data length between $\tilde{\mathbf{X}}$ and $Y$. We found that the overall results were quite robust with respect to sampling frequency ratio and window size: the estimated lag-CC profiles remained consistent, and the shape looked qualitatively similar (***Supplementary Fig. 3***). The window size affected the shape of the lag-CC profile, as an increasing window size decreased the temporal resolution of the lag-CC profile. Meanwhile, all CC statistics computed from surrogate data were significantly smaller.

**Bivariate GC systems with nonlinear coupling.** We further asked the question: how does the nonlinearity affect our estimation outcome in terms of identifiability? Specifically, we considered two variables that are Granger causally modulated by specific nonlinearity. We constructed two different types of nonlinear GC: one through phase-amplitude coupling (PAC), and the other through nonlinear amplitude coupling.

For PAC nonlinearity, we first generated raw signals $X_0$ and $Y$ using a linear bidirectional causal system as **Fig. 2a**. We then produced a modulated signal $X$ that was coupled with the phase of $X_0$, yielding a nonlinear coupling between $X$ and $Y$ (**Methods** and **Fig. 3a**). Upon down-sampling of $Y$, we found that xcorr (**Fig. 3b**) or MF-TFCCA correlating complex-valued STFT of $X$ with $Y$ (**Fig. 3c**) failed to discover the nonlinear causal relationship. In contrast, MF-TFCCA correlating the STFT amplitude of $X$ with $Y$ could uncover the nonlinear GC relationship (**Fig. 3d**). The reason for such result discrepancy is that the causal information was contained in the spectral power of $X$ in the PAC system but not in its fluctuations. The periodic nature of $X$ and the random phases in the complex-valued STFT lead to a reduced lag-CC profile. In contrast, the real-valued STFT considers only the amplitude component of the signal spectrum and is capable of recovering the true causality in the PAC system. Additionally, the CCA coefficients (in absolute value) at both positive and negative lags showed peaks in the HF range where $X$ stored the causal information of $Y$ (**Fig. 3e**).

For nonlinear amplitude coupling, we first generated raw signals $X_0$ and $Y$ using a linear bidirectional causal VAR system (***Supplementary Fig. 4a***) and then produced an amplitude-modulated signal $X$ by applying a nonlinear memoryless function of $X_0$ (such as a sigmoid function or a periodic cosine function; ***Supplementary Fig. 4b,c***; **Methods**). To detect such nonlinear causal relationship, we similarly applied MF-TFCCA correlating the complex-valued STFT $(X)$ with $Y$. Consequently, the lag-CC profile successfully recovered nonlinear GC relationship, which was significantly greater than the CC statistics of the surrogate data. As the amplitude modulation only enhances or suppresses the signals without altering the causal structure, MF-TFCCA with complex-valued STFT successfully revealed the causal relationship. On the opposite direction, when we applied nonlinear amplitude modulation to $Y$, and the results were similar (***Supplementary Fig. 4d,e***). Together, these results suggest that MF-TFCCA was capable of detecting nonlinear GC coupling between MF time series.

**Trivariate systems with chain and parallel GC systems.** To generalize the bivariate system to a trivariate system, we first considered the chain system and assumed that $X_1 \rightarrow X_2 \rightarrow Y$ with the known ground truth SGC profiles: $X_1 \rightarrow X_2|Y$ and $X_2 \rightarrow Y|X_1$ (**Fig. 4a**). More specifically, $X_1$ Granger causes $X_2$ at two driving frequencies (5 Hz and 80 Hz), and $X_2$ drives $Y$ at one frequency (50 Hz), with the MF setting as $F_{x_1} = F_{x_2} =$ 200 Hz and $F_y =$ 40 Hz (**Fig. 4b**). Direct xcorr analysis suggested an asymmetric information flow between $X_1$ and $Y$ (**Fig. 4c**). Provided that we only had access to $X_1$ and $Y$ from a partially observed system, the inferred SGC would suggest a direct causality between $X_1$ and $Y$ because of the latent variable $X_2$; however, the SGC inferred from the fully observed system would suggest negligible $X_1 \to Y|X_2$ information flow. In comparison, MF-TFCCA was also able to distinguish these two cases, where the partial CC (i.e., $X_1 \to Y|X_2$) was statistically insignificant (**Fig. 4d,e**).

Next, we considered the other forms of trivariate chain system $X \to Y_1 \to Y_2$, with the MF setting as $F_x =$ 200 Hz and $F_{y_1} = F_{y_2} =$ 40 Hz (***Supplementary Fig. 5a*** and **Methods**). We computed the STFT for both $X$ and $Y_1$ (***Supplementary Fig. 5b,c***). In this example, without considering $Y_1$, the lag-CC profile identified the directed information flow $X \to Y_2$ (***Supplementary Fig. 5d***); even partializing out the intermediate LF variable $Y_1$, the lag-CC profile still produced a similar result, yielding a false positive (***Supplementary Fig. 5e***). Alternatively, we could consider partializing out the STFT spectrum of $Y_1$ and repeated the analysis procedure. As a result, the lag-CC magnitude reduced but the estimate remained to be a false positive due to the information loss of down-sampling of $Y_1$.

Furthermore, we considered a trivariate parallel system $X_1 \to Y$ and $X_2 \to Y$ with a similar setting (***Supplementary Fig. 6a-c***), with the MF setting as $F_{x_1} = F_{x_2} =$ 200 Hz and $F_y =$ 40 Hz. The xcorr profile suggested a directed information $X_1 \to Y$ (***Supplementary Fig. 6c***). Additionally, the partial CC had the same estimate of standard CC (***Supplementary Fig. 6d,e***).

**Two-species Logistic model and coupled Rössler-Lorenz system.** Next, we validated our MF-TFCCA method with a two-dimensional deterministic nonlinear system: two-species Logistic model[30] (Ma et al., 2014), where the two generated time series can be either unidirectional ($X \to Y$, ***Supplementary Fig. 7a***) or bidirectional ($X \rightleftharpoons Y$, ***Supplementary Fig. 7e***) depending on the model parameters (**Methods**). As comparison, we also computed the standard GC between $X$ and $Y$ using the evenly-sampled time series (***Supplementary Fig. 7b,c,f,g***). We further down-sampled $Y$ and selected the sampling frequency ratio $\frac{F_x}{F_y}$ as 5:1. In both uni- and bidirectional cases, MF-TFCCA succeeded in recovering the correct directionality of information flow in the presence of nonlinearity (***Supplementary Fig. 7d,h***).

We further considered a nonlinear coupled system[30] (Ma et al., 2014), where 3-dimensional Rössler system $\boldsymbol{X} = \{x_1, x_2, x_3\}$ drives the 3-dimensional Lorenz system $\boldsymbol{Y} = \{y_1, y_2, y_3\}$, where both systems are deterministic chaotic and coupled through a unidirectional coupling term from $x_2$ to $y_2$ (**Methods**). The ground truth causality of the system is shown in ***Supplementary Fig. 8a.*** The goal was to infer the causality relationship between $\boldsymbol{X}$ and down-sampled $\boldsymbol{Y}$ by a factor of 5 (***Supplementary Fig. 8b***). As a comparison, we first computed the pairwise SGC between using the evenly-sampled time series (***Supplementary Fig. 8c,d***), and found significant information flow in many pairwise directions, including $x_2 \to y_2$, $x_2 \to y_3$ and $y_1 \to x_3, y_2 \to x_3, y_2 \to x_2, y_2 \to x_1$ directions. Noticeably, the SGC strength was greater in $y_2 \to x_2$ than $x_2 \to y_2$ direction. Additionally, some false-positive casual relationships were reported. In the MF time series setting, we applied MF-TFCCA to infer the relationship between $x_i$ $(i = 1,2,3)$ and $y_j$ $(j = 1,2,3)$ using real-valued version of STFT($x_i$) (***Supplementary Fig. 8e***).

In general, we found strong information flow in $y_1 \rightarrow x_3, y_2 \rightarrow x_3, y_3 \rightarrow x_3, y_3 \rightarrow x_2$ and $y_2 \rightarrow x_2, y_2 \rightarrow x_1$ directions. Notably, false-positive causal relationships were also detected from $\boldsymbol{Y}$ to $\boldsymbol{X}$. This result might be due to (i) the high periodicity of the chaotic systems, which could mislead the linear method to infer a false causality direction; (ii) the high correlation between the variables of the chaotic systems, such as between $x_1$ and $x_2$ and between $y_1$ and $y_2$. Together, these results suggest that the causality coupled through a highly nonlinear system may not be captured by the linear method.

**Benchmark comparison**

We further made systematic benchmark comparisons on GC inference (qualitatively) and computational efficiency (quantitatively) between our proposed MF-TFCCA and the state-of-the-art MF-VAR method. These results are summarized in **Supplementary Table 2** and ***Supplementary Fig. 9***.

First, more frequently, MF-VAR produced false-positive results in several systems. For example, MF-VAR mistakenly detected bidirectional causality in a truly unidirectional $Y \rightarrow X$ system, and mistakenly detected $X_1 \rightarrow Y|X_2$ in a fully observed $X_1 \rightarrow X_2 \rightarrow Y$ chain system. It even mistakenly detected an unreal bidirectional causality between $X_1$ and $X_2$ in a parallel system. In contrast, MF-TFCCA detected less false-positives. Moreover, MF-VAR failed to detect GC relationship in the presence of nonlinear coupling.

Second, MF-VAR can only detect GC in the time domain, whereas MF-TFCCA is capable of uncovering the driving frequency (i.e., SGC). When there are multiple driving frequencies present in the system, we can also assess the relative contribution of each driving frequency using the "filter-one-frequency-out" strategy.

More importantly, MF-TFCCA showed a greater advantage in computational efficiency compared to MF-VAR. In dealing with multiple-trial data structure, MF-TFCCA maintained an approximately linear time complexity, while the computational time of the MF-VAR inference method increased substantially (slower than quadratically) even in the simplest VAR(1) setting. In the general VAR(r) setting with a larger order *r*, the computational bottleneck is much worse for MF-VAR; whereas MF-TFCCA is nonparametric and independent of the VAR model order. Moreover, MF-TFCCA is efficient when computing across a broad range of time lags, whereas MF-VAR can only handle one lag each time. For instance, it took only a few minutes for MF-TFCCA to analyze a 100-trial trivariate system across a broad range of lags. In contrast, MF-VAR required around an hour to analyze a simple 2-trial trivariate system with two HF variables and one LF variable at only one lag. Therefore, it is nearly infeasible for MF-VAR to detect GC relationship with an uncertain lag or detect bidirectional causality with different lags (the null result on the bidirectional system in **Fig. 2** serves as an example).

Finally, we conducted a benchmark comparison on a real-world neuroscience dataset using concurrent local field potential (LFP) recordings of the rat primary somatosensory cortex (S1) and anterior cingulate cortex (ACC) during thermal pain experiments (**Fig. 5a; Methods**).

Consistent with our prior investigation[31] (Guo et al., 2020), we identified bidirectional SGC between the S1 and ACC in chronic pain-treated rats, where GC was prominent at theta (4-8 Hz) and gamma (50-70 Hz) bands at both S1→ACC and ACC→S1 directions derived from a bivariate VAR(15) model (**Fig. 5b**). Treating that as a "ground truth", we down-sampled the original sampling rate of LFP signals from one brain region by 4 folds (from 200 Hz to 50 Hz) and compared MF-TFCCA and MF-VAR results with the ground truth. Since our method is nonparametric, we also compared the parametric models with different orders: MF-VAR(1) and MF-VAR(5).

We first down-sampled S1 LFP signals and kept ACC LFP signals intact. We found that MF-TFCCA was able to identify bidirectional GC, whereas MF-VAR only identified unidirectional GC (**Fig. 5c**). MF-TFCCA was more computationally efficient than MF-VAR (by about 1-2 order of magnitude) at various settings (**Fig. 5d**). From the lag-CC curve (**Fig. 5e**), we inferred a bidirectional information flow. The CCA coefficients (absolute values) at both directions revealed two peaks, suggesting two driving frequencies (one at theta and another at gamma band). To further recover the dominant driving frequency, we systematically conducted the "filter-one-frequency-out" analysis and found a greater CC contribution at the theta-band driving frequency from its relative GC gains (**Fig. 5f**). Notably, both down-sampling (up to 50 Hz) and short-duration trials (1 second) and produced limited frequency resolution, preventing us to recover the gamma-band driving frequency. In contrast, the MF-VAR models were incapable of conducting similar analyses. At last, we reversed the direction by down-sampling ACC LFP signals and repeated the comparison, and the conclusion remained similar (**Fig. 5g-j**).

Together, these extensive benchmark comparisons pointed to the advantages of our proposed method compared to the state-of-the-art MF-VAR method. MF-TFCCA not only can correctly identify bidirectional spectral information flow, but also can uncover the driving frequency or quantify the relative contribution among multiple driving frequencies.

**Real-world finance and weather data**

To validate our method to real-world finance data, we studied public dataset that includes the year-to-year (YOY) growth rates of monthly Consumer Price Index (CPI), monthly OK WTI spot oil price (OIL) and quarterly real Gross Domestic Product (GDP) in the US (**Fig. 6a**). The CPI and OIL shares the same sampling rate, and standard VAR method detects GC from OIL to CPI (**Fig. 6b**). We then used TF-CCA to estimate the relationship between OIL and GDP and the relationship between CPI and GDP. The two lag-CC profiles revealed the causality from OIL to GDP (**Fig. 6c**), and a weaker causality from CPI to GDP (**Fig. 6d**). Next, we used partial CCA to assess the effect of CPI on the relationship between OIL and GDP, and found that causality relationship between OIL and GDP diminished (**Fig. 6e**). In conclusion, MF-TFCCA and GC methods reveal a similar causal chain from OIL to CPI to GDP. It is worth noting that MF-TFCCA can be used on a large range of lags without heavy computational burden, making it possible to detect potential long-term driving forces in the data.

Next, we applied MF-TFCCA to real-world climate data, which consisted of monthly and annual precipitation and temperature time series collected in Central Park, New York City from January 1869 to May 2024 (**Fig. 7a**). The goal here is to infer the causal relationship between monthly precipitation and monthly temperature, monthly precipitation and yearly temperature, and monthly temperature and yearly precipitation. Interestingly, there was only weak or non-significant GC between rainfall and temperature after removing the linear trend in the time series (**Fig. 7b**). By examining the MF time series, however, MF-TFCCA discovered a directed information flow between monthly rainfall and yearly temperature time series (**Fig. 7c**), but not between monthly temperature and yearly rainfall time series (**Fig. 7d**), suggesting that the detection result of true causal relationship of two random variables may depend on their relative temporal resolution.

**Concurrent hippocampal-neocortical recordings**

Advances in multi-region multimodal neural recordings (e.g., concurrent EEG-fMRI, or concurrent electrophysiology and calcium imaging, or concurrent electrophysiology and multifiber photometry) have enabled us to assess functional interactions between brain areas. Identifying directed information flow between upstream and downstream structures of the brain may help reveal state-dependent neural mechanisms.

We considered concurrent hippocampal-neocortical recordings that were publicly available[32] (Abadchi et al., 2020), which consisted of mouse hippocampal CA1 electrophysiological activity such as LFPs (<300 Hz) and multiunit activity (MUA, >300 Hz), as well as widefield calcium imaging (WFCI) activity of the restrosplenial cortex (RSC), primary visual cortex (V1), and forelimb primary somatosensory cortex (FLS1) (**Fig. 8a**).

It has been well known that the hippocampal MUA often increases during ripple events, especially in the UP or UP-to-DOWN state[33] (Buzsaki, 2015). First, we examined the relationship between hippocampal MUA and RSC activity, as the RSC is the closest downstream structure to the hippocampus. The hippocampal MUA was sampled at 2000 Hz, whereas the RSC activity extracted from GCaMP was sampled at 100 Hz, resulting a sampling frequency ratio 20:1 (**Fig. 8b**). Based on the previously detected hippocampal ripple events during non-rapid-eye-movement (NREM) sleep (n=753), we conducted STFT on the filtered hippocampal ripple band (100-250 Hz) activity using a 40-ms moving window with 30-ms overlapping window. Comparing the lag-CC profile and peak timing before, during and after the detected hippocampal ripple events (**Fig. 8c**, *i-iii*), we detected a bidirectional information flow between the RSC and hippocampus (HPC), with a slightly stronger strength in the RSC→HPC direction before and during hippocampal ripples. Additionally, the CC was significantly greater than the surrogate data (***Supplementary Fig. 10a***) and multiple driving frequencies were discovered (***Supplementary Fig. 10b***). The CC strength was the highest after the ripple events (**Fig. 8c**-*iv*), suggesting a possible cortico-hippocampal-cortical information loop covering before, during, and after ripple events. Among all tested trials, we sorted their CC scores (**Fig. 8d**) and separated the top 25% and bottom 25% trials (**Fig. 8e**). We further examined the trial averages of RSC activity and found the top 25% CC trials had higher amplitude (**Fig. 8f**). Next, we repeated the analysis for V1 and FLS1 and compared their lag-CC profiles during ripples. Similar to the bidirectional (but stronger RSC→HPC) information flow, we also found a bidirectional (but stronger V1→HPC) V1-HPC information flow but with a weaker strength than RSC-HPC (**Fig. 8g**). In contrast, the strength between FLS1 and HPC was the weakest (nearly indistinguishable from the surrogate data, ***Supplementary Fig. 10c***). Therefore, the hippocampal-cortical information flow during ripple events varies according to the downstream cortical structures, which is consistent with the finding based on large-scale electrophysiological findings[34] (Nitzan et al., 2022). We also repeated the analysis between the hippocampal LFP and RSC activity and found similar trends (***Supplementary Fig. 10d***).

## DISCUSSION

Identifying GC among MF time series has been studied in the field of statistics and econometrics[11,13,15,17] (Anderson et al., 2016; Ghysels et al., 2016; Tang et al., 2019; Shojaie and Fox, 2022). However, the current literature is limited to the discussion on the time domain or linearly coupled systems. In many physical and biological systems, directed information flow is associated with oscillatory frequency and it is more informative to interpret GC in the frequency domain and identify the driving frequency for the unidirectional or bidirectional GC system. Here we propose a new analysis pipeline to infer directed spectral information flow between MF time series. Our proposed MF-TFCCA is a nonparametric method for analyzing MF time series, which distinguishes itself from other parametric structural VAR methods in several aspects. First, parametric models require a data stationarity assumption, and often encounter problems in estimation biases (due to a limited sample size), model mismatch or identifiability; in contrast, MF-TFCCA requires no strong statistical assumption and is computationally efficient. Second, MF-TFCCA can not only recover the direction of information flow, but also reveal uni- or bidirectional driving frequencies and their relative strengths. Third, MF-TFCCA can potentially detect a wide range of nonlinear couplings between MF time series. Additionally, it can detect GC among more than two time series based on partialization of CCA. Therefore, our proposed method can serve as a hypothesis testing strategy for SGC during exploratory data analysis.

In MF-TFCCA estimation procedure, we convert the HF signal into a time-frequency representation before running CCA with the LF signals. Our choice of STFT as the time-frequency representation can be modified based on preference. Other types of time-frequency representations can be also considered, such as Wigner-Ville distribution (WVD), continuous wavelet spectrum (CWT), Stockwell-transformation (ST)[35] (Stockwell et al., 1996), short-time linear canonical transform (STLCT)[36] (Wei and Hu, 2021), and model-based parametric spectrum[37] (Ba et al., 2014). Specifically, WVD has several advantages over the STFT in that it can produce a high degree of resolution in both time and frequency, especially for non-stationary signals, and can distinguish between closely spaced frequency components. However, computation of STFT is more efficient due to its $O(N\log N)$ complexity. To characterize potentially nonlinear interactions between random variables, it is also possible to generalize CCA to nonlinear CCA or combine CCA with different preprocessing[25,38](Zhuang et al., 2020; Wang et al., 2020). Importantly, MF-TFCCA is nonparametric and is highly computationally efficient compared to other parametric methods such as MF-VAR.

We have validated our method using extensive computer simulations (**Supplementary Table 1**) and compared it with the standard MF-VAR method on benchmark experiments (**Supplementary Table 2**). In nearly all tested examples, MF-TFCCA was capable of identifying true directed information flow even in the presence of nonlinear coupling. However, various factors may contribute to false positive detection. Specifically, the effectiveness of our method is influenced by multiple factors, such as signal duration, driving frequency (in relation to the sampling frequency) and driving strength. For instance, constrained by the low sampling rate of LF signals, it requires that spectral causality is preserved with the LF range. Therefore, our method will perform better when the driving frequency is also low, or when the interference of other signals is at HF range. In the case of chain system, our method can somewhat mediate the intermediate variable using partial CCA, but is still prone to detect the false positive depending on the role of intermediate variable. In the case of complex nonlinear systems with highly periodic and correlated variables (e.g., coupled Rössler-Lorenz system), MF-TFCCA also identifies some false-positives of causal relationship. A systematic investigation of other nonlinear CCA methods for reducing the false-positive detection is beyond the scope of current paper and will be the subject of future investigation. Common challenges remain for detecting the causal relationship, such as strong autocorrelation, time delays or periodicity, and nonlinearity[39](Runge et al., 2019). Combing our method with parametric approaches to detrend and reduce the autocorrelation in the time series could be a potential solution. Our method can be easily integrated with the existing causal inference framework for causal hypothesis testing.

We have also validated our method in a wide range of real-world data. The use of CCA for identifying inter-area brain communications from neural signals has a long history[25] (Zhuang et al., 2020). Our proposed MF-TFCCA has replicated the GC results between the S1 and ACC LFP signals during pain experiments (Guo et al., 2020). Furthermore, MF-TFCCA can be viewed as a multimodal CCA method, which can analyze mixed-modal neural signals, such as concurrent electrophysiology and calcium imaging[40,41] (Wei et al., 2020; Patel et al., 2020), or concurrent EEG and fMRI[42] (Correa et al., 2010). Bidirectional hippocampal-neocortical communications are well known for sleep-dependent memory consolidation especially around hippocampal SWRs[33,43,44] (Buzsaki, 2015; Sirota et al., 2003; Pedrosa et al., 2022). When assessing hippocampal MUA and neocortical calcium imaging activity, we found a bidirectional yet asymmetric information flow with a stronger strength in the cortico-hippocampal direction. The degree of information flow from cortical areas to the hippocampus varied during ripple events, with the highest strength in the intermediate hippocampal downstream structure (RSC→HPC), and the weaker strength in sensory cortical areas (V1→HPC and FLS1→HPC). Additionally, we found differences in the RSC→HPC information flow before, during, and after hippocampal ripple events, where the cortico-hippocampal information was stronger before and during ripples. Concurrent mouse RSC and hippocampal LFP recordings during sleep have shown that the RSC displays a pre-ripple activation associated with slow and fast oscillations, and putative RSC inhibitory and excitatory neurons increase and decrease firing activities, respectively immediately after ripples[45] (Opalka et al., 2020). Recent rodent data have also shown the complexity of bidirectional hippocampal-neocortical communications during UP-DOWN states in NREM sleep[46] (Feliciano-Ramos et al., 2023), and their couplings are brain-state dependent[47] (Nitzan et al., 2020). It should be emphasized that the concept of Granger causality needs to be cautioned in self-organized systems such as the brain, where causes in such systems are often circular or multidirectional. Finally, the successes of MF-TFCCA in detecting directed causality in finance and weather data further demonstrated its broader applications and ability to discover weak directed GC relationship.

In general, inferring statistical dependency between MF variables presents a challenge for many well-studied estimation problems for evenly sampled data, including GC estimation[13] (Ghysels et al., 2016), regression and forecasting[48,49] (Ghysels et al., 2006; Bai et al., 2013), and stochastic control[50] (Sinopoli et al., 2004). The information of directed Granger causality may provide an intuitive strategy to forecast low-frequency variables (e.g., weekly) at higher frequencies (e.g., daily). In the context of inferring generalized SGC, compared to the standard MF-VAR method, our proposed analysis framework provides an exploratory and computationally efficient nonparametric approach to quantify directed information flow of MF time series in the presence of complex and nonlinear interactions. In the future, we expect to see more investigations to new explorations in physical and biological sciences.

## METHODS

**Canonical correlation analysis (CCA).** Without the loss of generality, Given two multivariate random variables $\mathbf{X} \in \mathbb{R}^m$ and $\mathbf{Y} \in \mathbb{R}^n$ (where the dimensionality of $m$ and $n$ may differ), CCA finds pairs of dimensions, such that the correlation between the projected activity onto these dimensions is maximally correlated: $\arg\max_{\mathbf{u},\mathbf{v}} \mathrm{corr}(\mathbf{Xu}, \mathbf{Yv})$, where the vectors $\mathbf{u}$ and $\mathbf{v}$ have respective dimensions. CCA can be estimated by solving a generalized eigenvalue decomposition problem:

$$\begin{bmatrix} \mathbf{0} & \mathbf{\Sigma}_{xy} \\ \mathbf{\Sigma}_{yx} & \mathbf{0} \end{bmatrix} \begin{bmatrix} \mathbf{u} \\ \mathbf{v} \end{bmatrix} = \rho \begin{bmatrix} \mathbf{\Sigma}_{xx} & \mathbf{0} \\ \mathbf{0} & \mathbf{\Sigma}_{yy} \end{bmatrix} \begin{bmatrix} \mathbf{u} \\ \mathbf{v} \end{bmatrix},$$

where $\mathbf{\Sigma}$ denotes the sample covariance matrix for specific random variables, and $\rho$ corresponds to the canonical correlation (CC). The largest squared CC, $\rho^2$, is used to measure the "set overlap." The CC also represents the association between the set of **X** and the set of **Y** after the within-set correlations have been removed, and CCA coefficients can be interpreted similar to the coefficients of principle components. CCA has also a natural link to multivariate regression. If the goal is to predict a target variable **Y** given an input vector **X** with a minimum squared error (MSE); let $\mathbf{Y} = \mathbf{a} + \mathbf{b}^T\mathbf{X}$, the optimal regression coefficient **b** is given by cross-correlation of two *whitened* variables $\mathbf{X}$ and $\mathbf{Y}$ [51](Jendoubi and Streimmer, 2019). A high value of coefficients in **b** indicates the relative importance of individual components in $\mathbf{X}$. In the special case when **Y** is univariate, the MSE reduces to variance-scaled coefficient of determination: $\mathbf{\Sigma}_{yy}\mathbf{\Sigma}_{yx}\mathbf{\Sigma}_{xx}^{-1}\mathbf{\Sigma}_{xy}$. To impose a sparsity constraint onto the CCA coefficients (or equivalently the regression coefficient **b**), regularized CCA algorithms have been designed[25] (Zhuang et al., 2020).

In time series analysis, traditional CCA assumes that two signals have zero lag. To generalize CCA to lagged CCA where two signals are temporally lagged, let $\Delta = t_1 - t_2$, we can reformulate the above equation as follows

$$\rho(\Delta) = \rho(t_1, t_2) = \max_{\mathbf{u},\mathbf{v}} \operatorname{corr}(\mathbf{X}_{t_1}\mathbf{u}, \mathbf{Y}_{t_2}\mathbf{v})$$

Partialization of CCA (also known as partial CCA) generalizes CCA and estimates a pair of linear projections onto a low dimensional space, where the correlation between two multivariate variables is maximized after eliminating the influence of a third variable[52,53] (Rao, 1969; Mukuta and Harada, 2014). Let $X$ and $Y$ be the target variables and $Z$ be the third variable, partial CCA is computed by solving a generalized eigenvalue decomposition problem as follows

$$\mathbf{\Sigma}_{xy|z}\mathbf{\Sigma}_{yy|z}^{-1}\mathbf{\Sigma}_{yx|z}\,\mathbf{u} = \rho\mathbf{\Sigma}_{xx|z}\mathbf{u}$$
$$\mathbf{\Sigma}_{yx|z}\mathbf{\Sigma}_{xx|z}^{-1}\mathbf{\Sigma}_{xy|z}\,\mathbf{v} = \rho\mathbf{\Sigma}_{yy|z}\mathbf{v}$$

where $\mathbf{\Sigma}_{\mu v|z} = \mathbf{\Sigma}_{\mu v} - \mathbf{\Sigma}_{\mu z}\mathbf{\Sigma}_{zz}^{-1}\mathbf{\Sigma}_{zv}$. In CCA or partial CCA, it is common to standardize all the variables.

### MF-TFCCA for three or more time series

To extend the analysis framework to the setting with multivariate $\mathbf{X}$ (HF process) and univariate $Y$ (LF process). For the sake of simple discussion, we assume that $\mathbf{X} = [X_1, X_2]$ is two-dimensional, where $X_1$ and $X_2$ denote two signals have the same temporal resolution. In this case, we apply STFT separately to $X_1$ and $X_2$, and then apply partial CCA to estimate the CC between $X_1$ and $Y$ conditional on $X_2$, or the CC between $X_2$ and $Y$ conditional on $X_1$.

In another setting with a univariate variable $X$ (HF process) and a multivariate variable $\mathbf{Y}$ (LF process). For the sake of simple discussion, we assume that $\mathbf{Y} = [Y_1, Y_2]$ is two-dimensional, where $Y_1$ and $Y_2$ denote two signals have the same temporal resolution. In this case, we apply STFT to $X$ and then apply partial CCA to assess the CC between $X$ and $Y_1$ conditional on $Y_2$, or the CC between $X$ and $Y_2$ conditional on $Y_1$.

### Lagged cross-correlation

We use the MATLAB function "xcorr" to compute time-lagged normalized cross-correlation function between two time series $x(t)$ and $y(t)$ in the same sampling rate. In the case of MF time series, we compute the correlation between $x(t)$ and $y(t)$ for different lags, with the temporal resolution of the lag determined by the larger sampling frequency.

### VAR and SGC estimate

For an arbitrary multivariate time series with the same sampling rate, we can model it using a *r*-order linear VAR(*r*) system for the demeaned multivariate random variable $\mathbf{\Theta} \in \mathbb{R}^n$

$$\boldsymbol{\Theta}_t = \sum_{\tau=1}^{r} \boldsymbol{A}_\tau \boldsymbol{\Theta}_{t-\tau} + \boldsymbol{e}_t$$

where $\mathbf{e}_t$ denotes *n*-dimensional white Gaussian noise, $\boldsymbol{A}_\tau$ denotes the *n*-by-*n* coefficient matrix for the $\tau$-th lag. If the VAR(*r*) process is stable, then all the roots of the reverse characteristic polynomial are bigger than 1 in terms of the Euclidean norm (i.e., outside the unit circle). When *r* is an even number, the VAR(*r*) can maximally produce $r/2$ oscillatory frequencies. Once $\{\boldsymbol{A}_\tau\}$ are known or fully identified, the SGC can be analytically computed[10] (Barnett and Seth, 2014). Note that when more than two variables are involved, the conditional SGC will be computed. In the special case of two variables, conditional GC and marginal GC are equivalent. The VAR(*r*) parameters can be identified from the least-squared estimation, following a model order selection for $r$. Between time series with the same sampling frequency, we can compute GC statistics in either time or frequency domain using an established MATLAB toolbox (www.sussex.ac.uk/sackler/mvgc/).

### MF-VAR and GC estimate

For multivariate time series with MF, we adopted the established MF-VAR model[13-15] (Ghysels et al., 2016; Gotz and Hecq, 2019; Anderson et al., 2016). Specifically, MF-VAR estimates GC by fitting a VAR(*r*) model of a stacked vector $\mathbf{\Theta}_t$:

$$\boldsymbol{\Theta}_t = \sum_{\tau=1}^{r} \boldsymbol{A}_\tau \boldsymbol{\Theta}_{t-\tau} + \boldsymbol{e}_t$$

$$\mathbf{\Theta}_t = \begin{bmatrix} x_H(t,1) \\ x_H(t,2) \\ \ldots \\ x_H(t,m) \\ x_L(t) \end{bmatrix}$$

where $x_H$ denotes the HF variable and $x_L$ denotes the LF variable. $m$ is the ratio between the high sampling frequency and the low sampling frequency. For implementation, we modified an existing MATLAB toolbox (http://www2.kobe-u.ac.jp/~motegi/Matlab_Codes.html) for computing single-trial MF-VAR into a multi-trial version.

### Surrogate data

To generate surrogate data for time series, we conducted the Fourier transform of time series, and randomly shuffled the phase while keeping the magnitude unchanged; we then conducted the inverse Fourier transform using the magnitude and randomized phase to reconstruct the surrogate time series. The procedure was repeated 100-500 times.

## Computer Simulations of VAR Processes

**Bivariate unidirectional systems.** The two random variables were first generated from a bivariate VAR(4) model

$$\begin{bmatrix} x_t \\ y_t \end{bmatrix} = \boldsymbol{A}_1 \begin{bmatrix} x_{t-1} \\ y_{t-1} \end{bmatrix} + \boldsymbol{A}_2 \begin{bmatrix} x_{t-2} \\ y_{t-2} \end{bmatrix} + \boldsymbol{A}_3 \begin{bmatrix} x_{t-3} \\ y_{t-3} \end{bmatrix} + \boldsymbol{A}_4 \begin{bmatrix} x_{t-4} \\ y_{t-4} \end{bmatrix} + \mathbf{e}_t$$

The diagonal elements of $\boldsymbol{A}_\tau = \begin{bmatrix} A_{11}(\tau) & A_{12}(\tau) \\ A_{21}(\tau) & A_{22}(\tau) \end{bmatrix}$ determine the oscillatory frequencies of $X$ and $Y$, and the off-diagonal elements determine causality between $X$ and $Y$. In generating the raw signals with a sampling rate of 200 Hz and a GC relationship from $X$ to $Y$, we assumed $X$ had two oscillatory frequencies ($f_1 = 80$ Hz and $f_2 = 4$ Hz), and $Y$ had two oscillatory frequencies ($f_3 = 15$ Hz and $f_4 = 2$ Hz), and there was SGC at 80 Hz and 4 Hz at $X \rightarrow Y$ direction (***Supplementary Fig. 2a***). The VAR(4) coefficient matrices were set as follows:

$$\boldsymbol{A}_1 = \begin{bmatrix} 2r_1\cos\theta_1 + 2r_2\cos\theta_2 & 0 \\ -0.4 & 2r_3\cos\theta_3 + 2r_4\cos\theta_4 \end{bmatrix}$$

$$\boldsymbol{A}_2 = \begin{bmatrix} -r_1^2 - r_2^2 - 4r_1r_2\cos\theta_1\cos\theta_2 & 0 \\ 0.7 & -r_3^2 - r_4^2 - 4r_3r_4\cos\theta_3\cos\theta_4 \end{bmatrix}$$

$$\boldsymbol{A}_3 = \begin{bmatrix} 2r_1r_2(r_1\cos\theta_1 + r_2\cos\theta_2) & 0 \\ -0.1 & 2r_3r_4(r_3\cos\theta_3 + r_4\cos\theta_4) \end{bmatrix}$$

$$\boldsymbol{A}_4 = \begin{bmatrix} -r_1^2r_2^2 & 0 \\ 0 & -r_3^2r_4^2 \end{bmatrix}$$

where $r_1 = 0.9, r_2 = 0.8, r_3 = 0.85, r_4 = 0.7$, and $\theta_1 = 2\pi f_i \Delta_t$ ($i = 1,2,3$) and $\Delta_t = 5$ ms denotes the sampling interval. Note that the coefficients $\boldsymbol{A}_{(1)21} = -0.4$ in $\boldsymbol{A}_1$, $\boldsymbol{A}_{(2)21} = 0.7$ in $\boldsymbol{A}_2$ and $\boldsymbol{A}_{(3)21} = -0.1$ in $\boldsymbol{A}_3$ control the degree of causality.

In the bivariate system $Y \rightarrow X$ (***Supplementary Fig. 2h***), we assumed that $X$ and $Y$ had the same oscillations as in the $X \rightarrow Y$ system. The VAR(4) coefficient matrices were set as follows:

$$\boldsymbol{A}_1 = \begin{bmatrix} 2r_1\cos\theta_1 + 2r_2\cos\theta_2 & 0.05 \\ 0 & 2r_3\cos\theta_3 + 2r_4\cos\theta_4 \end{bmatrix}$$

$$\boldsymbol{A}_2 = \begin{bmatrix} -r_1^2 - r_2^2 - 4r_1r_2\cos\theta_1\cos\theta_2 & -0.05 \\ 0 & -r_3^2 - r_4^2 - 4r_3r_4\cos\theta_3\cos\theta_4 \end{bmatrix}$$

$$\boldsymbol{A}_3 = \begin{bmatrix} 2r_1r_2(r_1\cos\theta_1 + r_2\cos\theta_2) & 0.1 \\ 0 & 2r_3r_4(r_3\cos\theta_3 + r_4\cos\theta_4) \end{bmatrix}$$

$$\boldsymbol{A}_4 = \begin{bmatrix} -r_1^2r_2^2 & 0 \\ 0 & -r_3^2r_4^2 \end{bmatrix}$$

Note that although the off-diagonal coefficients were smaller than the $X \rightarrow Y$ coefficients, the driving force in the $Y \rightarrow X$ system could be higher than that in the $X \rightarrow Y$ system, since the causality is also related with the spectrum of the transmitter[54] (Stokes and Purdon, 2017).

In computer simulations, we down-sampled $Y$ by a factor *K* (e.g., *K*=5, $F_y = 40$ Hz) while keeping the sampling rate of $X$ intact ($F_x = 200$ Hz). To avoid aliasing in down-sampling, we conducted a low-pass filtering before decimation (see **Supplementary Note**). To compute the STFT of $X$, we chose a window size of 200 ms and the overlap for the moving window as 200-25=175 ms (i.e., window size minus sampling interval of $Y$); consequently, the matrix $\widetilde{\mathbf{X}}$ has the same length in the time-axis as time series vector $Y$.

**Bivariate bidirectional system.** We generated two random variables with a bidirectional causal relationship $X \leftrightarrow Y$ using a bivariate VAR(41) model. To discriminate the two driving directions, the VAR order was large as 41 to create the lags for each direction (**Fig. 2a**). We assumed that $X$ had an oscillatory frequency at $f_1 = 4$ Hz and $Y$ had an oscillatory frequency at $f_2 = 15$ Hz. Both oscillatory frequencies were in the low frequency band in order to maintain the bidirectional causality after down-sampling of $Y$. $X$ was sampled at 200 Hz and $Y$ was sampled at 40 Hz. The first two VAR(41) coefficient matrices control the oscillations of $X$ and $Y$:

$$\boldsymbol{A}_1 = \begin{bmatrix} 2r_1\cos\theta_1 & 0 \\ 0 & 2r_2\cos\theta_2 \end{bmatrix}$$

$$\boldsymbol{A}_2 = \begin{bmatrix} -r_1^2 & 0 \\ 0 & -r_2^2 \end{bmatrix}$$

where $r_1 = 0.8$ and $r_2 = 0.9$. Additionally, coefficient matrices $\boldsymbol{A}_{21}$ to $\boldsymbol{A}_{23}$ control the causality $X \rightarrow Y$ with a lag of $20/f_s = 0.1$ s. $\boldsymbol{A}_{(21)21} = -0.175, \boldsymbol{A}_{(22)21} = 0.35, \boldsymbol{A}_{(23)21} = -0.175$. Coefficient matrices $\boldsymbol{A}_{31}$ to $\boldsymbol{A}_{23}$ control the causality $Y \rightarrow X$ with a lag of $30/f_s = 0.15$ s. The elements $\boldsymbol{A}_{(30+i)12}$ $(i = 1,2, \ldots, 11)$ were set to be the $i$ element of a 10th-order band-pass filter: $[0.000, 0.001, -0.014, -0.039, 0.026, 0.098, 0.026, -0.039, -0.014, 0.001, 0.000]$. Other elements were all set as 0.

**Trivariate chain system.** We generated three random variables using a trivariate VAR(4) model

$$\begin{bmatrix} x_{1,t} \\ x_{2,t} \\ y_t \end{bmatrix} = \boldsymbol{A}_1 \begin{bmatrix} x_{1,t-1} \\ x_{2,t-1} \\ y_{t-1} \end{bmatrix} + \boldsymbol{A}_2 \begin{bmatrix} x_{1,t-2} \\ x_{2,t-2} \\ y_{t-2} \end{bmatrix} + \boldsymbol{A}_3 \begin{bmatrix} x_{1,t-3} \\ x_{2,t-3} \\ y_{t-3} \end{bmatrix} + \boldsymbol{A}_4 \begin{bmatrix} x_{1,t-4} \\ x_{2,t-4} \\ y_{t-4} \end{bmatrix} + \mathbf{e}_t$$

In the chain system $X_1 \rightarrow X_2 \rightarrow Y$, we assumed that $X_1$ had two resonant frequencies at $f_1 = 80$ Hz and $f_2 = 5$ Hz, $X_2$ had an oscillatory frequency at $f_3 = 15$ Hz, and $Y$ had two oscillatory frequencies at $f_4 = 10$ Hz and $f_5 = 2$ Hz (**Fig. 4a**). $X_1$ and $X_2$ were sampled at 200 Hz, whereas $Y$ was sampled at 40 Hz. Further, we assumed that $X_1 \rightarrow X_2$ strongly and $X_2 \rightarrow Y$ moderately using the following VAR(4) setup:

$$\boldsymbol{A}_1 = \begin{bmatrix} 2r_1\cos\theta_1 + 2r_2\cos\theta_2 & 0 & 0 \\ -0.7 & 2r_3\cos\theta_3 & 0 \\ 0 & -0.3 & 2r_4\cos\theta_4 + 2r_5\cos\theta_5 \end{bmatrix}$$

$$\boldsymbol{A}_2 = \begin{bmatrix} -r_1^2 - r_2^2 - 4r_1r_2\cos\theta_1\cos\theta_2 & 0 & 0 \\ 1.5 & -r_3^2 & 0 \\ 0 & 0.4 & -r_4^2 - r_5^2 - 4r_4r_5\cos\theta_4\cos\theta_5 \end{bmatrix}$$

$$\boldsymbol{A}_3 = \begin{bmatrix} 2r_1r_2(r_1\cos\theta_1 + r_2\cos\theta_2) & 0 & 0 \\ 1 & 0 & 0 \\ 0 & -0.3 & 2r_4r_5(r_4\cos\theta_4 + r_5\cos\theta_5) \end{bmatrix}$$

$$\boldsymbol{A}_4 = \begin{bmatrix} -r_1^2 r_2^2 & 0 & 0 \\ 0 & 0 & 0 \\ 0 & 0 & -r_4^2 r_5^2 \end{bmatrix}$$

where $r_1 = 0.95, r_2 = 0.7, r_3 = 0.85, r_4 = 0.85, r_5 = 0.7$. When all three variables were known, the system are completely observed; if only $X_1$ and $Y$ can be accessed, the system is a partially observed chain system.

For the chain system $X \to Y_1 \to Y_2$, time-series data were generated using the same VAR model (***Supplementary Fig. 5a***), except for the intermediate variable sampled at 40 Hz instead of 200 Hz.

**Trivariate parallel system.** In the parallel system, we assumed that $X_1 \to Y$ weakly and $X_2 \to Y$ strongly using the following new VAR(4) coefficients (***Supplementary Fig. 6a***):

$$\boldsymbol{A}_1 = \begin{bmatrix} 2r_1\cos\theta_1 + 2r_2\cos\theta_2 & 0 & 0 \\ 0 & 2r_3\cos\theta_3 & 0 \\ -0.4 & -0.8 & 2r_4\cos\theta_4 + 2r_5\cos\theta_5 \end{bmatrix}$$

$$\boldsymbol{A}_2 = \begin{bmatrix} -r_1^2 - r_2^2 - 4r_1r_2\cos\theta_1\cos\theta_2 & 0 & 0 \\ 0 & -r_3^2 & 0 \\ 0.7 & 1.5 & -r_4^2 - r_5^2 - 4r_4r_5\cos\theta_4\cos\theta_5 \end{bmatrix}$$

$$\boldsymbol{A}_3 = \begin{bmatrix} 2r_1r_2(r_1\cos\theta_1 + r_2\cos\theta_2) & 0 & 0 \\ 0 & 0 & 0 \\ -0.1 & -1 & 2r_4r_5(r_4\cos\theta_4 + r_5\cos\theta_5) \end{bmatrix}$$

$$\boldsymbol{A}_4 = \begin{bmatrix} -r_1^2 r_2^2 & 0 & 0 \\ 0 & 0 & 0 \\ 0 & 0 & -r_4^2 r_5^2 \end{bmatrix}$$

where $r_1 = 0.95, r_2 = 0.7, r_3 = 0.85, r_4 = 0.85, r_5 = 0.7$. $X_1, X_2$ and $Y$ had the same oscillatory frequencies as in the chain system. In the case of parallel system, the results of fully observed and partially observed systems are similar, since the missing variable will not affect the GC between the other two variables. However, the missing variable ($X_1$ or $X_2$) may interfere with specific frequencies of $Y$, making it more challenging to identify SGC.

**Trivariate VAR(3) model.** We also studied a generic trivariate VAR system (***Supplementary Fig. 1d***) in the previous study[54] (Stokes and Purdon, 2017)

$$\begin{bmatrix} x_{1,t} \\ x_{2,t} \\ x_{3,t} \end{bmatrix} = \boldsymbol{A}_1 \begin{bmatrix} x_{1,t-1} \\ x_{2,t-1} \\ x_{3,t-1} \end{bmatrix} + \boldsymbol{A}_2 \begin{bmatrix} x_{1,t-2} \\ x_{2,t-2} \\ x_{3,t-2} \end{bmatrix} + \boldsymbol{A}_3 \begin{bmatrix} x_{1,t-3} \\ x_{2,t-3} \\ x_{3,t-3} \end{bmatrix} + \mathbf{e}_t$$

where $\boldsymbol{A}_1 = \begin{bmatrix} 2r_1\cos\theta_1 & 0 & 0 \\ -0.356 & 2r_2\cos\theta_2 & 0 \\ 0 & -0.3098 & 2r_3\cos\theta_3 \end{bmatrix}, \boldsymbol{A}_2 = \begin{bmatrix} -r_1^2 & 0 & 0 \\ 0.7136 & -r_2^2 & 0 \\ 0 & 0.5 & -r_3^2 \end{bmatrix}, \boldsymbol{A}_3 = \begin{bmatrix} 0 & 0 & 0 \\ -0.356 & 0 & 0 \\ 0 & -0.3098 & 0 \end{bmatrix}$.

The parameters are set as follows: $r_1 = 0.9, r_2 = 0.7, r_3 = 0.8$, $f_1 = 40$ Hz, $f_2 = 10$ Hz, $f_3 = 50$ Hz, $\theta_i = 2\pi f_i \Delta_t$, and $\Delta_t = 1/120$ ms (i.e., sampling frequency 120 Hz).

**Phase amplitude coupling (PAC) system.** We used a two-step procedure to generate a pair of signals with PAC. In step 1, we generated raw signals $X_0$ and $Y$ using a linear bidirectional causal system as described in **Fig. 2a**. In step 2, we viewed $X_0$ as the phase signal with a resonating frequency at $f_0 = 4$ Hz and generated a modulated signal with its amplitude coupling with the phase of $X_0$ as follows[55] (Munia and Kviyente, 2019):

$$X_0(t) = K_{f_0}(t)\sin(2\pi f_0 t)$$
$$X_a(t) = \left(K_{f_0}(t)\sin(2\pi f_0 t) + const\right)\sin(2\pi f_a t) = (X_0(t) + const)\sin(2\pi f_a t)$$

With $const = \left|\min\left(X_0(t)\right)\right|$ and $f_a = 90$ Hz; which thereby yielded a nonlinear PAC coupling between $X = X_a$ and $Y$ (***Supplementary Fig. 3a***).

### Two-species Logistic Model

The nonlinear two-species Logistic model was described in the following difference equations[30] (Ma et al., 2014),

$$X(t+1) = X(t)\left[r_x - r_x X(t) - \gamma_{xy} Y(t)\right]$$
$$Y(t+1) = Y(t)\left[r_y - r_y Y(t) - \gamma_{yx} X(t)\right]$$

where $r_x = 3.7$ and $r_y = 3.8$ are self-regulation parameters; $\gamma_{xy}$ and $\gamma_{yx}$ are two coupling constants. In the unidirectional case (***Supplementary Fig. 7a***), we set the coupling constants as $\gamma_{xy} = 0$ and $\gamma_{yx} = 0.32$; in the bidirectional case (***Supplementary Fig. 7e***), we set $\gamma_{xy} = 0.02$ and $\gamma_{yx} = 0.1$. We used uniformly distributed random numbers in [0,1] as the system's initial conditions and discarded the initial 100 time points after the transient dynamics. To create MF time series, we subsequently down-sampled $Y$ by a factor of 5.

### Coupled Rössler-Lorenz System

The Rössler attractor and Lorenz attractor are two deterministic chaotic systems. To create a complex nonlinear coupled system, we used the same computer simulation setting as REF[30] (Ma et al., 2014), in which a 3-dimensional nonlinear Rössler system ***X*** drives a 3-dimensional nonlinear Lorenz system ***Y*** (***Supplementary Fig. 8a***), which is described by the following differential equations

$$\dot{x}_1 = -\alpha(x_2 + x_3)$$
$$\dot{x}_2 = \alpha(x_1 + 0.2x_2)$$
$$\dot{x}_3 = \alpha(0.2 + x_3(x_1 - 5.7))$$
$$\dot{y}_1 = 10(-y_1 + y_2)$$
$$\dot{y}_2 = 28y_1 - y_2 - y_1 y_3 + Cx_2^2$$
$$\dot{y}_3 = y_1 y_2 - 8/3 y_3$$

where $\alpha = 6$ is a timescale constant and $C = 2$ denotes the strength of unidirectional coupling. The initial values of the system were randomly chosen and the original sampling frequency for both systems was 100 Hz. To create MF time series, we subsequently down-sampled ***Y***= $\{y_1, y_2, y_3\}$ by a factor of 5.

### Finance Data

The financial time series consists of the year-over-year (YOY) growth rates of monthly Consumer Price Index (CPI), monthly OK WTI spot oil price (OIL) and quarterly real Gross Domestic Product (GDP) in the United States. The CPI, OIL and GDP data are published by the U.S. Department of Labor, the Energy Information Administration, and the Bureau of Economic Analysis, respectively. These data were collected over the course of 30 years (ranging from January 1987 to January 2016), consisting of 360 monthly observations and 120 quarterly observations. The monthly CPI and OIL time series share the same sampling rate, whereas quarterly GDP time series has the lower sampling rate. The YOY growth rates for each value were computed as follow:

$$\text{YOY Growth} = \left(\frac{\text{currect period value}}{\text{prior period value}} - 1\right) \times 100\%$$

### Climate Data

The climate data consist of average monthly and annual precipitation and temperatures (starting from January 1869 to May 2024) at the Central Park of New York City, which are publicly available at the National Weather Service (https://www.weather.gov/media/okx/Climate/CentralPark/monthlyannualprecip.pdf and https://www.weather.gov/media/okx/Climate/CentralPark/monthlyannualtemp.pdf). The time series consist of 1865 monthly observations and 155 yearly observations. Given this dataset, the estimation goal is to infer the causal relationships between monthly precipitation and monthly temperature, monthly precipitation and yearly temperature, and monthly temperature and yearly precipitation time series.

### Concurrent Rat Cortical LFP Recordings

In a prior study[31] (Guo et al., 2020), noxious pain stimuli were used for freely exploring male adult Sprague-Dale rats (250-300 g) in a plastic chamber of size 38×20×25 $cm^3$ on top of a mesh table. In thermal pain experiments, a blue (473 nm diode-pumped solid-state) laser with 250 mW intensity was delivered to the rat's hindpaw. The laser stimulation was delivered in repeated trials (25-40) during 30–45 min. The stimulation was terminated by animal's paw withdrawal. To produce chronic inflammatory pain, 0.075–0.1 ml of *Complete Freund's adjuvant* (CFA) was suspended in an oil-saline (1:1) emulsion, and injected subcutaneously into the plantar aspect of the hindpaw opposite to the paw that was stimulated by a blue laser. Namely, only a unilateral inflammation was induced, and nociceptive stimuli were delivered to the opposite paw of the injured foot. All experimental studies were performed in accordance with the National Institutes of Health (NIH) *Guide for the Care and Use of Laboratory Animals* to ensure minimal animal use and discomfort, and were approved by the New York University Grossman School of Medicine Institutional Animal Care and Use Committee (IACUC).

We used silicon probes (NeuroNexus) with 3D printed drive to record multi-channel (up to 64 channels) neural activities from the rat ACC and S1 areas simultaneously. The probe implant was on the contralateral side of the paw that received noxious stimulation. The Plexon (Dallas, TX) data acquisition system was used to record *in vivo* extracellular neural signals at a sampling rate of 40 kHz. Local field potential (LFP) signals were further preprocessed and down-sampled to 200 Hz. Details have been presented elsewhere (Guo et al., 2020)[31]. For the purpose of benchmark comparison, we selected some preprocessed LFP data from a single CFA rat during 250 mW laser stimulation trials. Each laser stimulation trial lasted 1 second. Data used in our analysis are also publicly available (https://github.com/QiqiXian/MF-TFCCA).

**Multimodal Mouse Hippocampal-Neocortical Recordings**

In a prior study[32] (Abadchi et al., 2020), concurrent hippocampal-neocortical recordings were collected from head-restrained mice during natural sleep and urethane anesthesia. Specifically, wide-field mesoscale neocortical activity was imaged based on optical imaging and voltage-sensitive dye (VSD) in the animal's right hemisphere, and in vivo electrophysiological recordings, including LFPs and multi-unit activity (MUA), were simultaneously measured in the animal's ipsilateral hippocampus. The VSD imaging employed a CCD camera and recorded 12-bit images at 100 Hz sampling rate. Sensory stimulation was used to determine the coordinates for the primary sensory areas, whereas the relative locations of associated areas were estimated using stereotaxic coordinates. The animal protocols were approved by the University of Lethbridge Animal Care Committee and were in accordance with guidelines set forth by the Canadian Council for Animal Care.

Muscle electromyogram (EMG) recordings and hippocampal delta-to-theta band power ratio to score the sleep state. VSD imaging signals were denoised, filtered and preprocessed to obtain the ΔF/F activity relative to the baseline. Hippocampal LFP was first down-sampled to 2 kHz, sharp-wave ripples (SWRs) were identified using ripple-band filter combined with an established threshold criterion. Condition on the detected hippocampal SWRs, that patterns of activity in neocortical regions were differentially modulated around hippocampal SWRs. Three regions of interest (ROIs) in neocortical areas, including the retrosplenial cortex (RSC), primary visual cortex (V1), and forelimb primary somatosensory cortex (FLS1), were identified and selected in our current analysis. Neural recordings were preprocessed in advance and are publicly available (https://doi.org/10.5061/dryad.qnk98sfbb). However, the original raw recordings were not available, limiting the scope of our analyses.

**Data Availability.** The real-world data in this study are publicly available. The data for computer simulations can be accessed via the shared software.

**Code Availability.**

The computer simulations and codes developed in this study are available in the GitHub repository (https://github.com/QiqiXian/MF-TFCCA).

**Acknowledgments.** This work was initiated during a summer research internship when Q.X. was an undergraduate student at the University of Science and Technology of China. A preliminary version of this work was presented in the IEEE ICASSP'24, Seoul, April 13-19, 2024. The work was supported by grants R01-MH118928, P50-MH132642, R01-NS121776, R01-MH139352 and RF1-DA056394 (Z.S.C.) from the US National institutes of Health.

**Author contributions.** Z.S.C. conceived and supervised experiments, developed the methods, interpreted the data, and wrote the paper. Q.X. developed the methods, performed experiments, analyzed and interpreted the data. Z.S.C. acquired funding.

**Competing interests**

The authors declare no competing interests.

**Supplementary information**

Supplementary note, Figs. S1-S10 and Tables S1-S2.

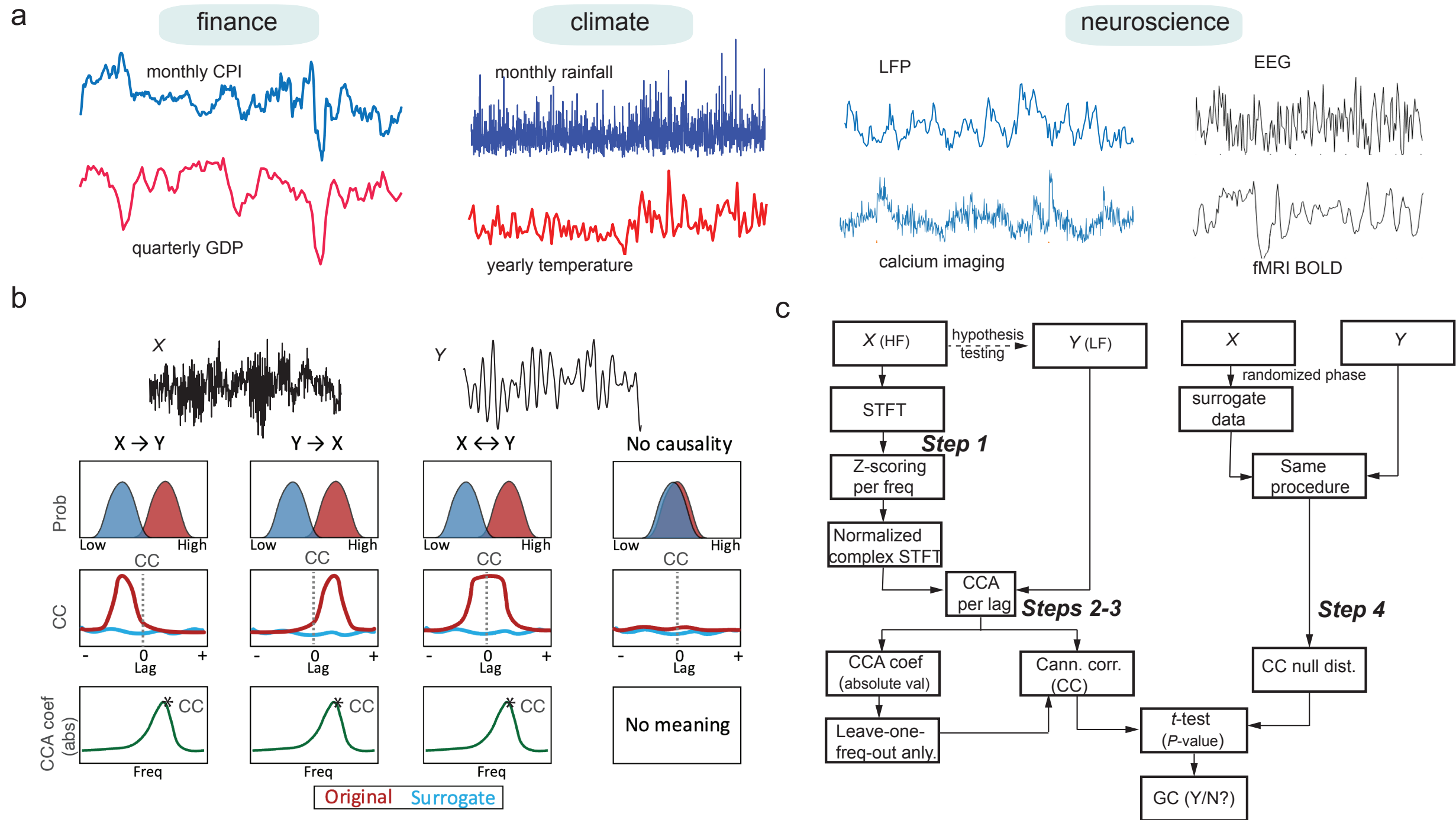

*(Insert)* **Figure 1. Illustration of mixed-frequency (MF) time series and estimation of directed spectral information.**

- **(a)** Examples of MF time series observed in finance, climate, and neuroscience applications.
- **(b)** Cartoon illustration of two subtask criteria to identify directed information flow between the HF signal $X$ and LF signal $Y$ (top row). The second row compares two distributions of canonical correlation (CC) of a specific information flow direction at a specific time lag derived from the original and surrogate data. The third row compares the lag-CC profiles of the original and surrogate data. The shape of lag-CC profile indicates direction of information flow: unidirectional, bidirectional, or non-significant. The fourth row examines the CCA coefficients (in absolute value) computed from MF-TFCCA. The peak (*) indicates the putative driving frequency.
- **(c)** Flowchart of the MF-TFCCA method (Steps 1 through 4).

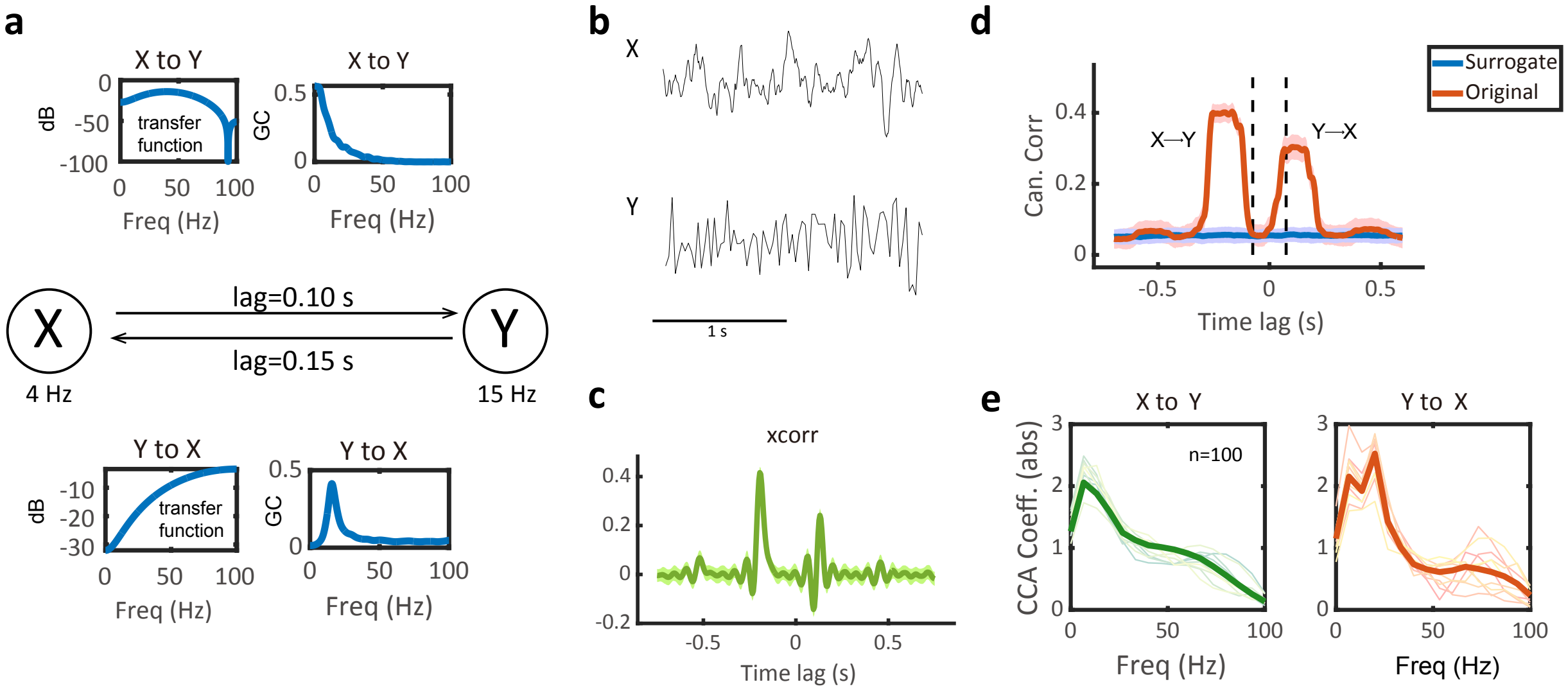

*(insert)* **Figure 2. Computer simulation results on a bivariate bidirectional Granger causality (GC) system.**

- **(a)** Random variables $X$ and $Y$ are bidirectionally causally related by a VAR(4) system. The transfer function profile and SGC ground truth are shown for both $X$ to $Y$ (top) and $Y$ to $X$ (bottom) directions. Both variables $X$ and $Y$ have oscillatory frequencies in the low-frequency band and are able to keep the causal relationship even after down-sampling $Y$.
- **(b)** Snapshots of bivariate time series $X(t)$ and $Y(t)$.
- **(c)** Lagged cross-correlation profile generated from xcorr(X,Y). The positive peak indicates that $X$ lags $Y$, and the negative peak indicates that $X$ leads $Y$.
- **(d)** Lag-CC profile generated from linear MF-TFCCA, which detects the bidirectional causal relationship and the time lags of two directions. Two vertical dashed lines denote the window size at both positive and negative lags (relative to origin) that define the temporal resolution to detect spectral information flow. Red trace denotes the CC estimate derived from the original data, and blue trace denotes the CC estimate derived from the surrogate data. Shade areas denote 95% confidence intervals (n=100). The significant CC profile located beyond the left dashed line suggests causal information flow from $X$ to $Y$, whereas significant CC profile located beyond the right dashed line suggests causal information flow from $Y$ to $X$.
- **(e)** The CCA coefficients (in absolute value) computed from MF-TFCCA in two directions. The CCA coefficients at the negative lag showed a peak at about 4 Hz, corresponding to the driving frequency of the information from $X$ to $Y$. The CCA coefficients at the positive lag showed a peak at about 15 Hz, corresponding to the driving frequency of the information from $Y$ to $X$, and a lower peak at 4 Hz. Solid trace denotes the average (n=100).

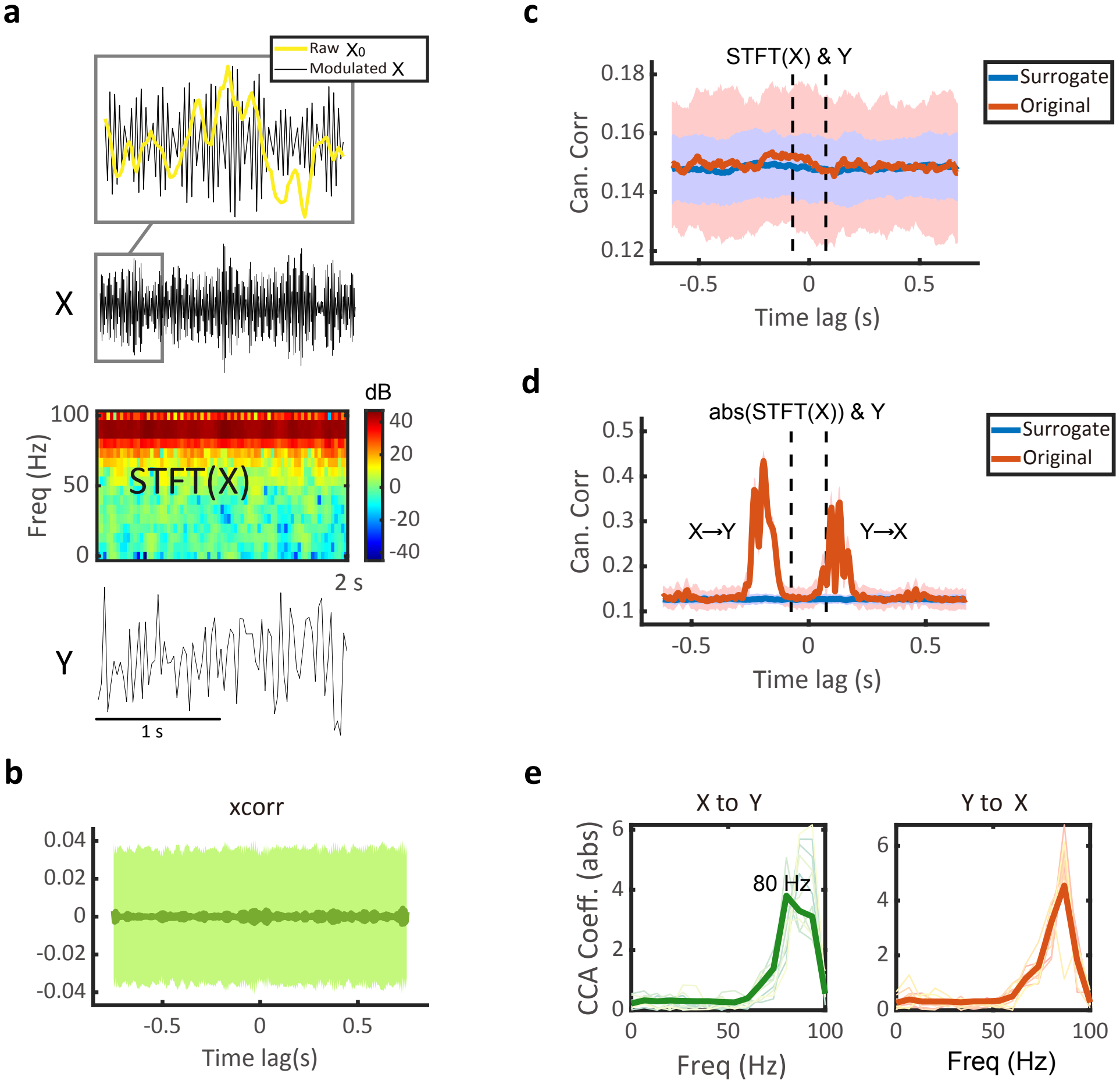

*(insert)* **Figure 3. Computer simulation results on a bivariate bidirectional Granger causality (GC) system with nonlinear phase-amplitude coupling (PAC).**

- **(a)** An illustration of the amplitude of the modulated signal $X$ (black) was coupled with the phase of the raw signal $X_0$ (yellow), and $X_0$ and $Y$ were generated by a bidirectional causal VAR(2) system ($X_0 \leftrightarrow Y$) similar to the setup in **Figure 2a**.
- **(b)** The lagged cross-correlation profile generated from xcorr(X,Y), which failed to capture any GC structure.
- **(c)** The lag-CC profile generated from MF-TFCCA based on the complex-valued STFT($X$) failed to detect GC relationship between $X$ and $Y$.
- **(d)** The lag-CC profile generated from MF-TFCCA based on the amplitude of STFT($X$), which successfully detected bidirectional GC relationship between $X$ and $Y$ at both positive and negative lags. The CC was computed with a 150-ms window.
- **(e)** The CCA coefficients (in absolute value) computed from MF-TFCCA at both positive (left) and negative (right) lags showed peaks in the high-frequency range where $X$ stored the causal information of $Y$. Solid trace denotes the average.

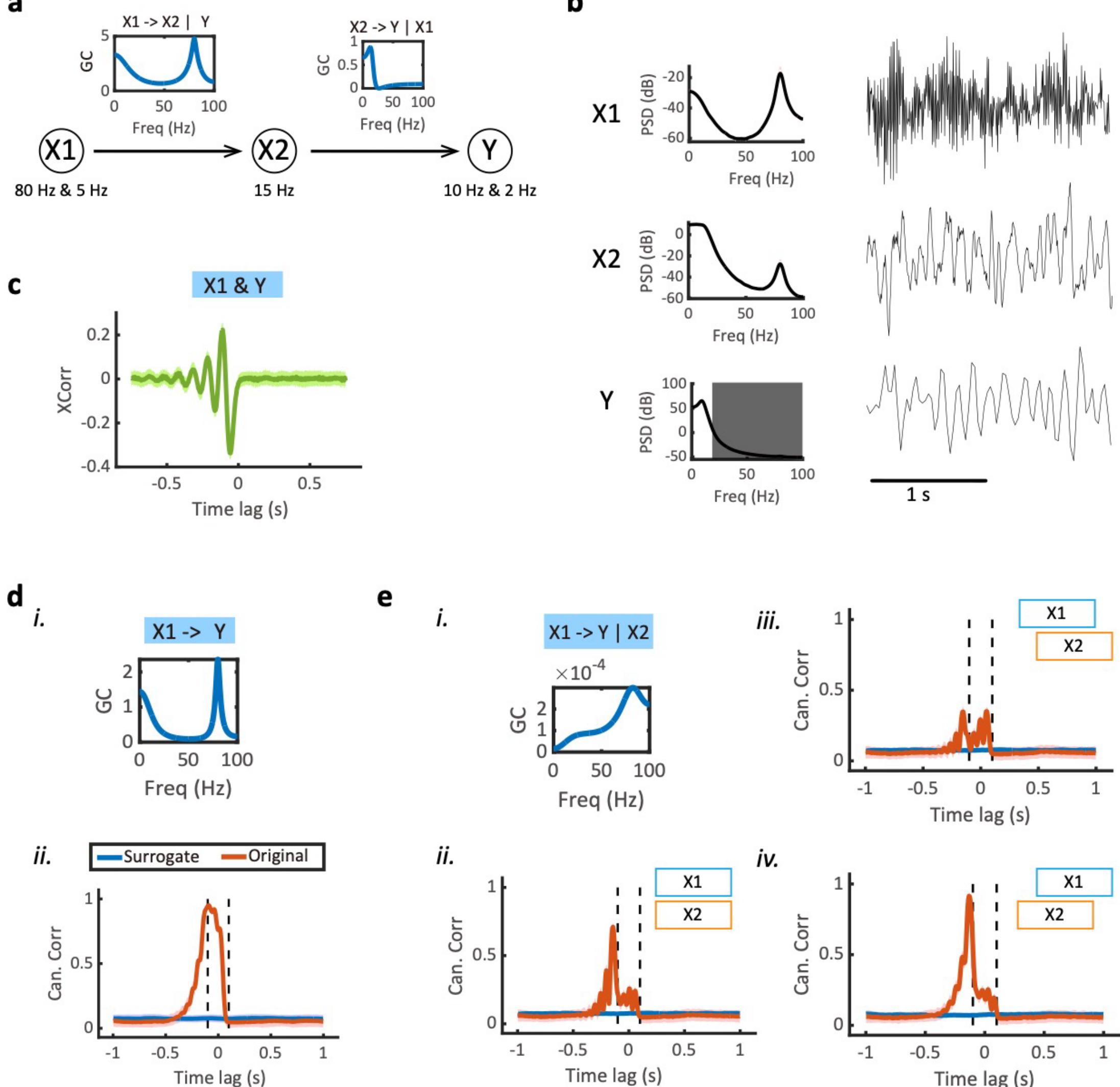

*(insert)* **Figure 4. Computer simulation results on trivariate Granger causality (GC) in a chain system.**

- **(a)** Illustration of a chain system $X_1 \rightarrow X_2 \rightarrow Y$ and two conditional SGC profiles $X_1 \rightarrow X_2|Y$ and $X_2 \rightarrow Y|X_1$. Note that $X_1$ drives $X_2$ at two frequencies, and $X_2$ drives Y at one frequency.
- **(b)** Snapshots of trivariate time series and their power spectral density (PSD). Here $X_1$ and $X_2$ have high sampling rates (200 Hz), whereas $Y$ has a low sampling rate (40 Hz). The shaded area denotes the filtered-out frequency band above the cut-off frequency (20 Hz) of down-sampled $Y$.
- **(c)** The lagged cross-correlation profile generated from xcorr($X_1$,Y).
- **(d)** The SGC profile $X_1 \rightarrow Y$ and the lag-CC profile inferred from MF-TFCCA. Red trace denotes the estimate derived from the original data. The significant CC profile located beyond the left dashed line suggests a causal information flow from $X_1$ to $Y$.
- ***(e)*** Similar to panel **d**, except for the conditional SGC profile $X_1 \rightarrow Y|X_2$ and lag-CC profile by partializing out $X_2$. (*i*) The conditional SGC value was statistically insignificant. (*ii-iv*) MF-TFCCA detected a directed information flow from $X_1$ to $Y$ (i.e., a false-positive), but the magnitude of the lag-CC reduced compared to panel **d**(*ii*). The exact shape of lag-CC curve also changed depending on relative time lag between $X_2$ and $X_1$. Since $X_1 \rightarrow X_2$, the CC estimate was the smallest when $X_2$ was leading $X_1$ (*iii*) and was the greatest when $X_2$ was lagging $X_1$ (*iv*). The relative lag of $X_1$ and $X_2$ is illustrated by the two-colored boxes in the figure legend.

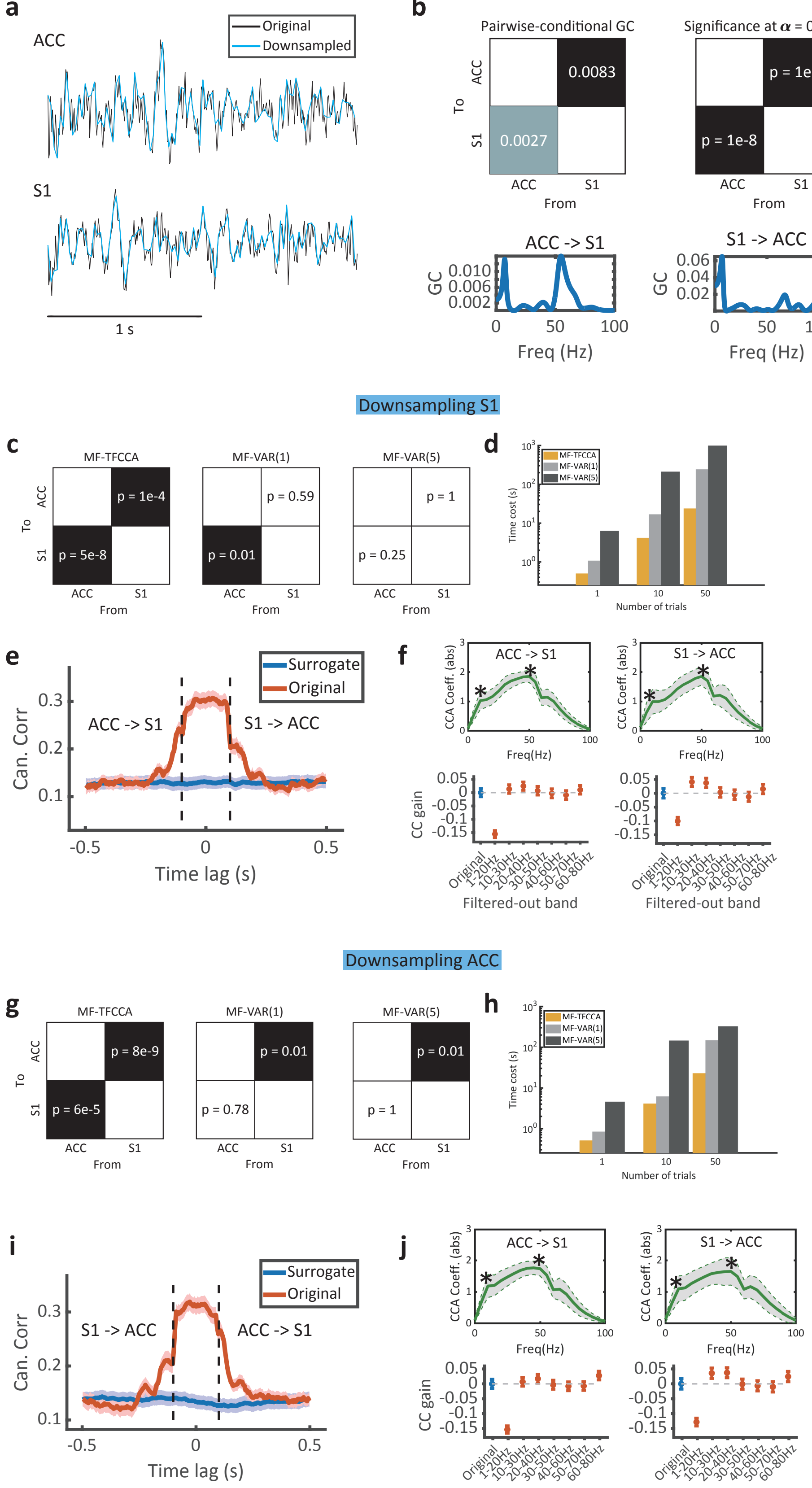

*(insert)* **Figure 5. Benchmark comparison between MF-TFCCA and MF-VAR for real-world neuroscience data.**

(**a**) Exemplar LFP traces from the rat ACC and S1.

(**b**) Standard GC test in both time and frequency domains for equally sampled ACC and S1 time series (sampling rate: 200 Hz). Bidirectional GC was inferred based on a VAR(15) model (n=59 trials, 1 s per trial): ACC drives S1 at around 7 Hz and 55 Hz, whereas S1 drives ACC at around 7 Hz and 65 Hz. P-values smaller than 0.05 indicate statistical significance.

(**c**) Mixed-frequency GC tests for ACC and S1 LFP signals. The S1 LFP signal was down-sampled by a factor of 4 (i.e., 50 Hz sampling rate). MF-TFCCA detected bidirectional causality, while MF-VAR(1) only detected unidirectional ACC→S1 causality, and MF-VAR(5) failed to detect any causality.

(**d**) Computational cost of the three models with respect to the number of trials. Noe that the plot is in log-log scale. The computational cost included computing 100 repetitions of surrogate data.

(**e**) The lag-CC profile generated from MF-TFCCA revealed the bidirectional information flow. The negative lag denotes ACC→S1 direction in information flow, where the positive lag denotes S1→ACC direction in information flow.

(**f**) *Top:* The CCA coefficients (absolute values) revealed two local peaks (marked by *) related to the driving frequencies. *Bottom*: The follow-up "filter-one-frequency-out" analysis recovered the dominant theta-band driving frequency (i.e., a large negative CC gain) at both ACC→S1 and S1→ACC directions. Error bar denotes SEM (n= 59 trials).

(**g**) Similar to panel **c**, but with the ACC LFP signals down-sampled by a factor of 4. MF-TFCCA detected bidirectional causality, while MF-VAR(1) and MF-VAR(5) detected unidirectional S1→ACC causality.

(**h-j**) Similar to panels **d-f**, except that the direction is swapped.

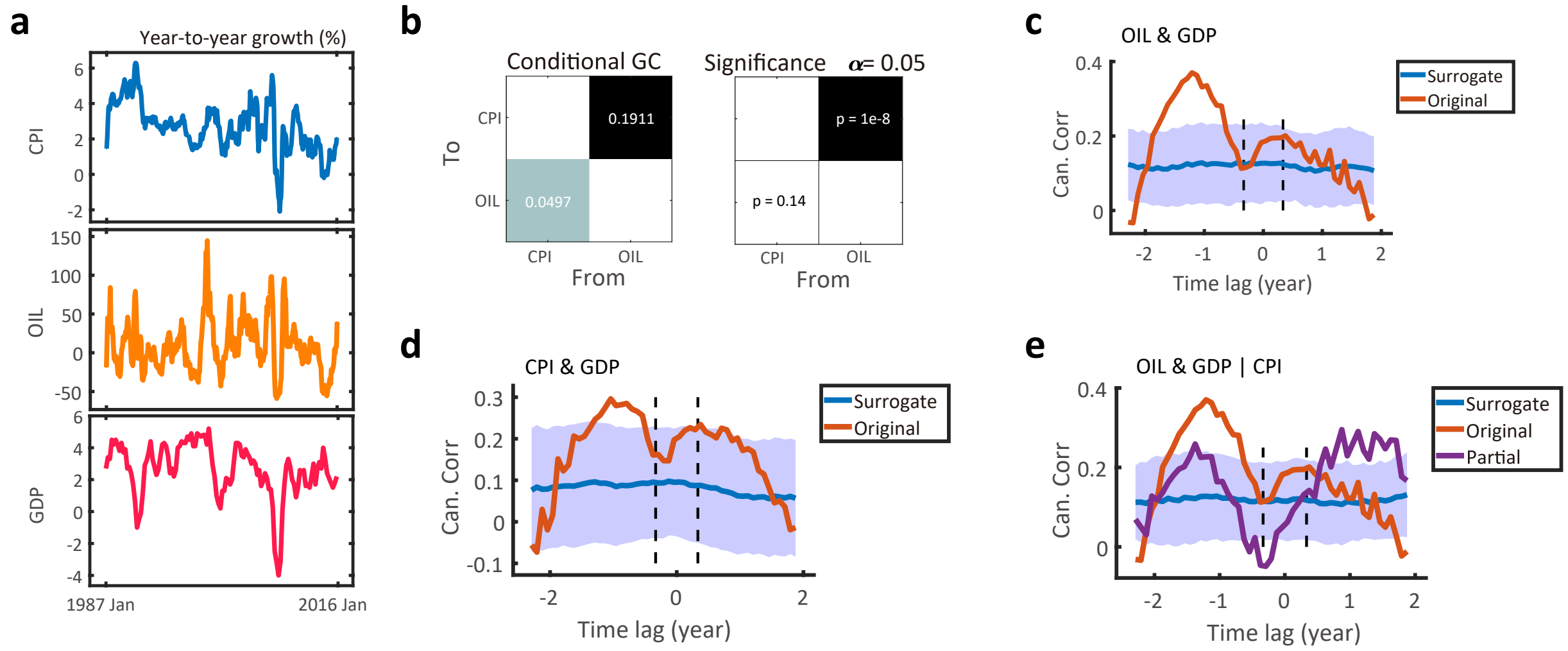

*(insert)* **Figure 6. Estimation results of directed information flow for real-world finance data.**

- **(a)** Year-to-year growth rates of CPI, OIL and GDP time series over 30 years (from January 1987 to January 2016). The monthly OIL and CPI time series have 360 time points, whereas quarterly GDP has 120 time points.
- **(b)** The standard GC test detected the causality from OIL to CPI.
- **(c)** The lag-CC profile generated from MF-TFCCA detected a clear causality from monthly OIL to quarterly GDP. The CC estimate (red trace) from the original data was computed using an 8-month window. Blue trace denotes the estimate derived from the surrogate data. Shade areas denote 95% confidence intervals (n=100).
- **(d)** The lag-CC profile generated from MF-TFCCA detected a weaker causality from monthly CPI to quarterly GDP, and non-significant causality from quarterly GDP to monthly CPI.
- **(e)** The lag-CC profile derived from partial CCA between monthly OIL and quarterly GDP conditional on monthly CPI showed that GC was statistically insignificant compared to the surrogate data, indicating that the causality from OIL to GDP was mediated by CPI.

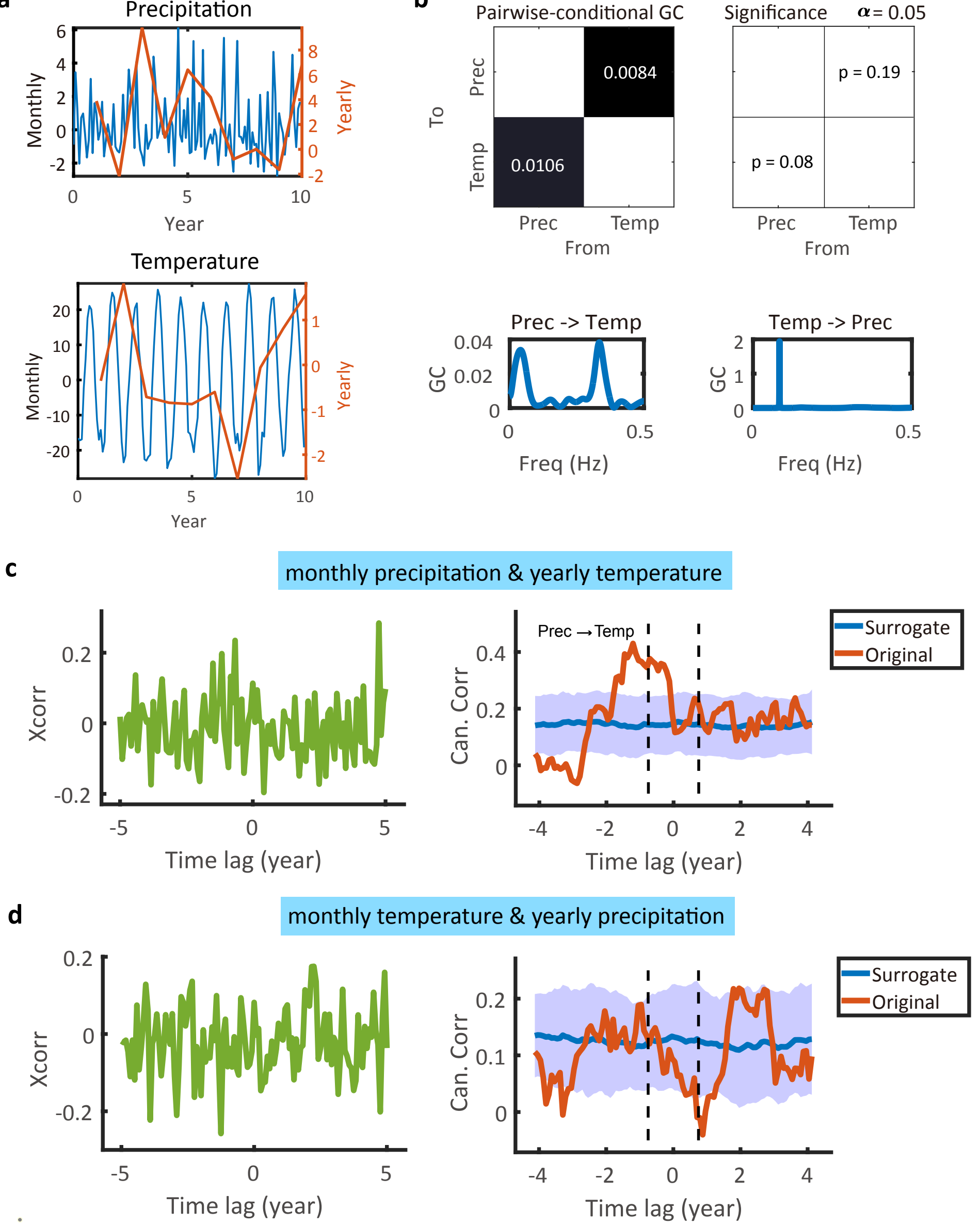

*(insert)* **Figure 7. Estimation results of directed information flow for real-world climate data.**

- **(a)** The standard GC Selected snapshot of monthly/yearly precipitation and monthly/yearly temperature time series.
- **(b)** Standard GC test in both time and frequency domains for equally sampled monthly precipitation and temperature time series. Weak GC was found in the precipitation→temperature direction.
- **(c)** Estimated lagged cross-correlation and lag-CC profile between monthly precipitation and yearly temperature time series. A significant directed information from precipitation to temperature was found.
- **(d)** Estimated lagged cross-correlation and lag-CC profile between monthly temperature and yearly precipitation time series. No significant directed information flow was identified.

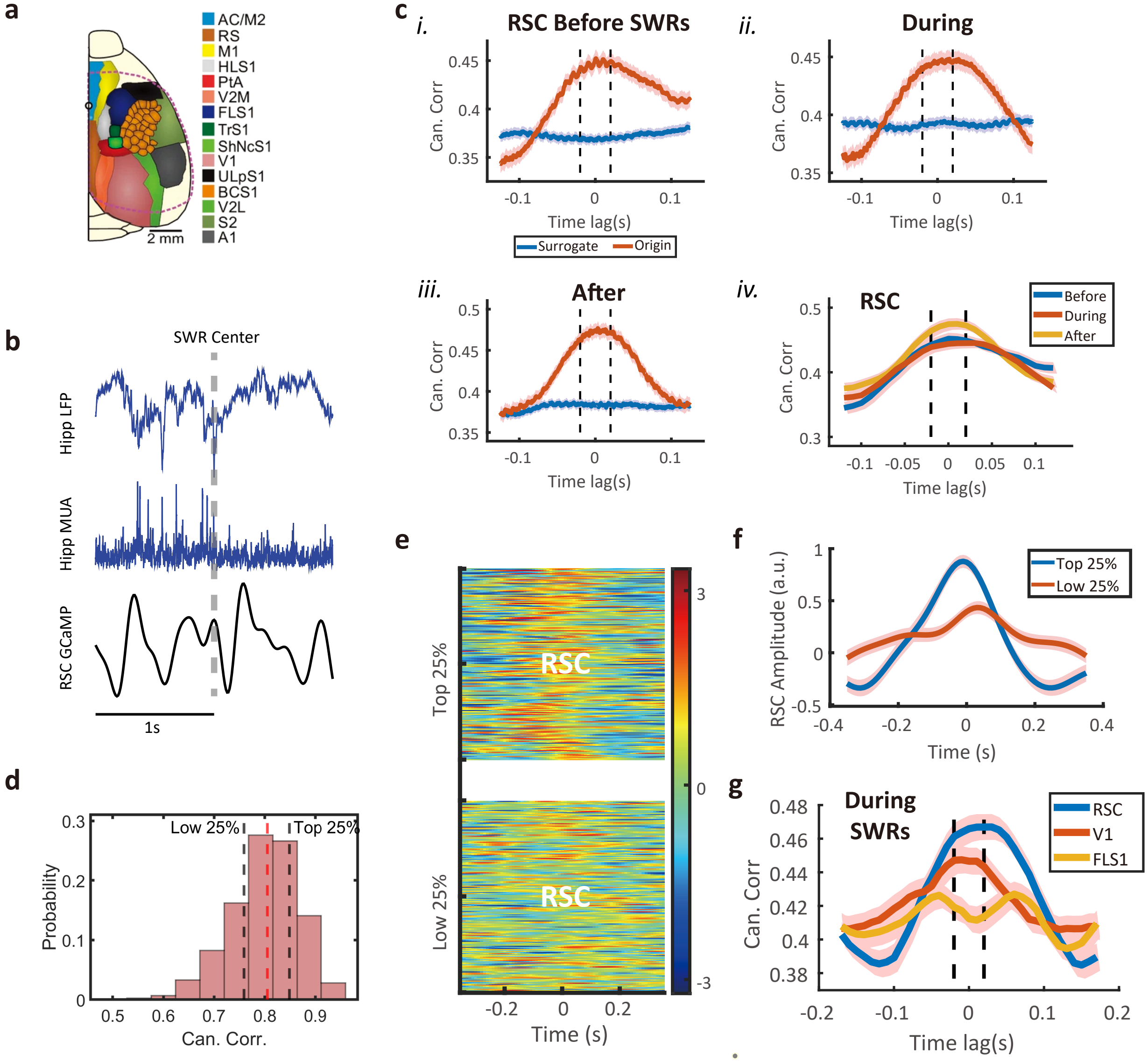

*(insert)* **Figure 8. Estimation results of rodent hippocampal-neocortical recordings.**

**(a)** Schematic of a cranial window for wide-field optical imaging of neocortical activity using voltage or glutamate probes. The voltage or glutamate signal was recorded from dorsal surface of the right neocortical hemisphere, containing the specified regions. Adapted from REF[31] (Abadchi et al., 2020; CC-BY license).

**(b)** An exemplar hippocampal raw LFP trace, hippocampal MUA, and RSC GCaMP.

**(c)** Comparison of lag-CC profiles between hippocampal MUA and RSC activity before (*i*), during (*ii*), and after (*iii*) hippocampal ripples. (*i-iii*) Red trace denotes the estimate derived from the original data, and blue trace denotes the estimate derived from the surrogate data. Shade areas denote 95% confidence intervals. A stronger directed information flow RSC→HPC was detected before and during hippocampal ripples, whereas the lag-CC curve became more symmetric after hippocampal ripples. (*iv*) Comparison of CC values between three conditions (*i-iii*).

**(d)** Distribution of CC between the RSC activity and hippocampal MUA.

**(e)** Heatmaps of RSC activity sorted according to the top 75% and bottom 25% of canonical correlation. Time 0 marks the onset of hippocampal ripples.

**(f)** The averaged amplitude of RSC activity from the two groups shown in panel **e**.

**(g)** Comparison of three lag-CC profiles between hippocampal MUA and RSC/V1/FLS1 activities during hippocampal ripples. The asymmetric shape suggests a stronger of information flow in RSC→HPC, followed by V1→HPC, followed by the weakest strength between FLS1 and HPC.

# Supplementary Information

## Inferring directed spectral information flow between mixed-frequency time series

Qiqi Xian and Zhe Sage Chen

### Supplementary Note

Given two raw random time series $x(t)$ and $y(t)$, we may compute their cross-correlation function $R_{xy}(t)$. In the frequency domain, the Fourier transform of the cross-correlation function is written as: $\mathcal{F}\{R_{xy}(t)\} = S_{xy}(f) = X(f)Y^*(f)$. Let's further assume that two raw signals are filtered by two filters $g$ and $h$, respectively; the two filtered signals are represented as: $\tilde{x}(t) = x(t) * g$ and $\tilde{y}(t) = y(t) * h$. Let $R_{\tilde{x}\tilde{y}}(t)$ denote the cross-correlation function between $\tilde{x}(t)$ and $\tilde{y}(t)$, then the Fourier transform of $R_{\tilde{x}\tilde{y}}(t)$ is written as $\mathcal{F}\{R_{\tilde{x}\tilde{y}}(t)\} = S_{\tilde{x}\tilde{y}}(f) = S_{xy}(f)G(f)H^*(f)$. If $x(t)$ and $y(t)$ are convolved with the same temporal low-pass filter, it may introduce interactions between instantaneous (i.e. zero-lag) and time-lagged relationships; in other words, there may be "leakage" of zero-lag correlation into time-lagged GC because of autocorrelation of $\tilde{x}(t)$ and $\tilde{y}(t)$ due to low-pass filtering. Therefore, the interpretation of results requires a careful examination of the observed signals. The standard VAR modeling can be modified to accommodate the instantaneous effect (e.g., Nuzzi et al., 2021; Desphande et al., 2010).

Down-sampling or subsampling requires a low-pass filter before decimation in the time domain, so that the sample frequency needs to be at least twice of the max frequency signal, otherwise the aliasing may emerge (i.e., high-frequency signal components will copy into the lower frequency band and be mistaken for lower frequencies). For instance, if $y(t)$ is represented by a discrete-time signal $y[n]$, with a down-sampling factor of *D*, we will obtain $y[Dn]$. In the frequency domain, the magnitude of Fourier spectrum of $y[Dn]$ is lowered by a factor of *D* compared to the magnitude of Fourier spectrum of $y[n]$. To avoid aliasing (especially when *D* is rather large), the cut-off frequency of the low-pass filter should be $0.5F_y$, where $F_y$ denotes the sampling frequency of $y[Dn]$. Therefore, this combined operation (low-pass filter plus down-sampling) can be viewed as a special case of low-pass time-varying filter, which is not invariant to temporal shift. Consequently, down-sampling has an impact on the time-lagged causality.

### Supplementary references

**Table S1. Summary of all computer simulation results.**

| Condition | Driving frequency | Direction of information flow | Figure |
|---|---|---|---|
| Bivariate system $\{X, Y\}$, $X(t)$ HF process, $Y(t)$ LF process | | | |
| $X \overset{f_x}{\to} Y$ (linear, unidirectional) | Recover $f_x$ | Identify $X \to Y$ from lag-CC | Fig. S2 |
| $X \overset{f_x}{\to} Y$ (linear, unidirectional) | Recover $f_y$ | Identify $Y \to X$ from lag-CC | Fig. S2 |
| $X \overset{f_x}{\to} Y$ and $Y \overset{f_y}{\to} X$ (linear, bidirectional, $f_x$>$f_y$) | Recover $f_x$ but not necessarily $f_y$ (depending on $F_y$) | Identify dominant direction (amplitude difference) | Fig. 2 |
| $X \overset{f_x}{\to} Y$ and $Y \overset{f_y}{\to} X$ (bidirectional, nonlinear PAC) | Recover $f_x$ and $f_y$ | Identify dominant direction, recover the modulated (but not original) driving frequency | Fig. 3 |
| $X \overset{f_x}{\to} Y$ and $Y \overset{f_y}{\to} X$ (bidirectional, sigmoidal amplitude coupling) | Recover $f_x$ and $f_y$ | Identify dominant direction and driving frequency | Fig. S4 |
| $X \overset{f_x}{\to} Y$ and $Y \overset{f_y}{\to} X$ (bidirectional, sinusoidal amplitude coupling) | Recover $f_x$ and $f_y$ | Identify dominant direction and driving frequency | Fig. S4 |
| Trivariate system $\{X_1, X_2, Y\}$ or $\{X, Y_1, Y_2\}$ | | | |
| $X_1 \to X_2 \to Y$ (linear, unidirectional, chain system | Recover $f_x$ | Misidentify $X_1 \to Y \mid X_2$ as putative GC pattern (lag-CC depends on the temporal shift of $X_2$) | Fig. 4 |
| $X \to Y_1 \to Y_2$ (linear, unidirectional, chain system) | Recover $f_x$ | Misidentify $X \to Y_2 \mid Y_1$ as putative GC pattern (lag-CC depends on the $Y_1$ term in the time or time-frequency domain) | Fig. S5 |
| $X_1 \to Y$ and $X_2 \to Y$ (linear, unidirectional, parallel system) | Recover $f_x$ | Identify $X_1 \to Y \mid X_2$ and $X_2 \to Y \mid X_1$, compared to $X_1 \to Y$ and $X_2 \to Y$ | Fig. S6 |
| two-species Logistic model $\{X, Y\}$ | | | |
| Nonlinear, uni- or bidirectional | n/a | Identify the directionality of information flow | Fig. S7 |
| coupled Rössler-Lorenz system $\{\boldsymbol{X}, \boldsymbol{Y}\}$ | | | |
| Nonlinear, unidirectional | n/a | Identify the directionality of information flow | Fig. S8 |

**Table S2. Summary results of qualitative comparison between MF-TFCCA and MF-VAR methods. In this table, the down-sampling factor is 5 and sampling frequency ratio is 5:1. False positives are marked in red.**

| Condition | MF-TFCCA detected information flow | MF-VAR detected information flow |
|---|---|---|
| $X \overset{f_x}{\rightarrow} Y$ (linear, unidirectional) | $X \rightarrow Y$<br>(Fig. S1) | $X \rightarrow Y$ |
| $Y \overset{f_y}{\rightarrow} X$ (linear, unidirectional) | $Y \rightarrow X$<br>(Fig. S1) | $X \rightarrow Y$ and $Y \rightarrow X$ |
| $X \overset{f_x}{\rightarrow} Y$ and $Y \overset{f_y}{\rightarrow} X$ (linear, bidirectional) | $X \rightarrow Y$ and $Y \rightarrow X$<br>(Fig. 2) | None |
| $X_1 \rightarrow X_2 \rightarrow Y$ (linear, unidirectional, chain system | Partially observed: $X_1 \rightarrow Y$<br>Fully observed: *reduced* $X_1 \dashrightarrow Y \vert X_2$ and $X_2 \rightarrow Y \vert X_1$ (Fig. 4) | Partially observed: $X_1 \rightarrow Y$<br>Fully observed: $X_1 \rightarrow Y \vert X_2$ and $X_2 \rightarrow Y \vert X_1$ |
| $X \rightarrow Y_1 \rightarrow Y_2$ (linear, unidirectional, chain system) | Partially observed: $X \rightarrow Y_2$<br>Fully observed: *reduced* $X \dashrightarrow Y_2 \vert Y_1$<br>(Fig. S5) | Partially observed: $X \rightarrow Y_2$<br>Fully observed: $X \rightarrow Y_1 \vert Y_2$ and $X \rightarrow Y_2 \vert Y_1$ |
| $X_1 \rightarrow Y$ and $X_2 \rightarrow Y$ (linear, unidirectional, parallel system) | Partially and fully observed:<br>$X_1 \rightarrow Y$ and $X_2 \rightarrow Y$<br>(Fig. S6) | Partially observed:<br>$X_1 \rightarrow Y$ and $X_2 \rightarrow Y$<br>Fully observed:<br>$X_1 \rightarrow Y$, $X_2 \rightarrow Y$ and $X_1 \leftrightarrow X_2$ |
| $X \overset{f_x}{\rightarrow} Y$ and $Y \overset{f_y}{\rightarrow} X$ (nonlinear PAC, bidirectional) | $X \rightarrow Y$ and $Y \rightarrow X$<br>(Fig. 3) | None |

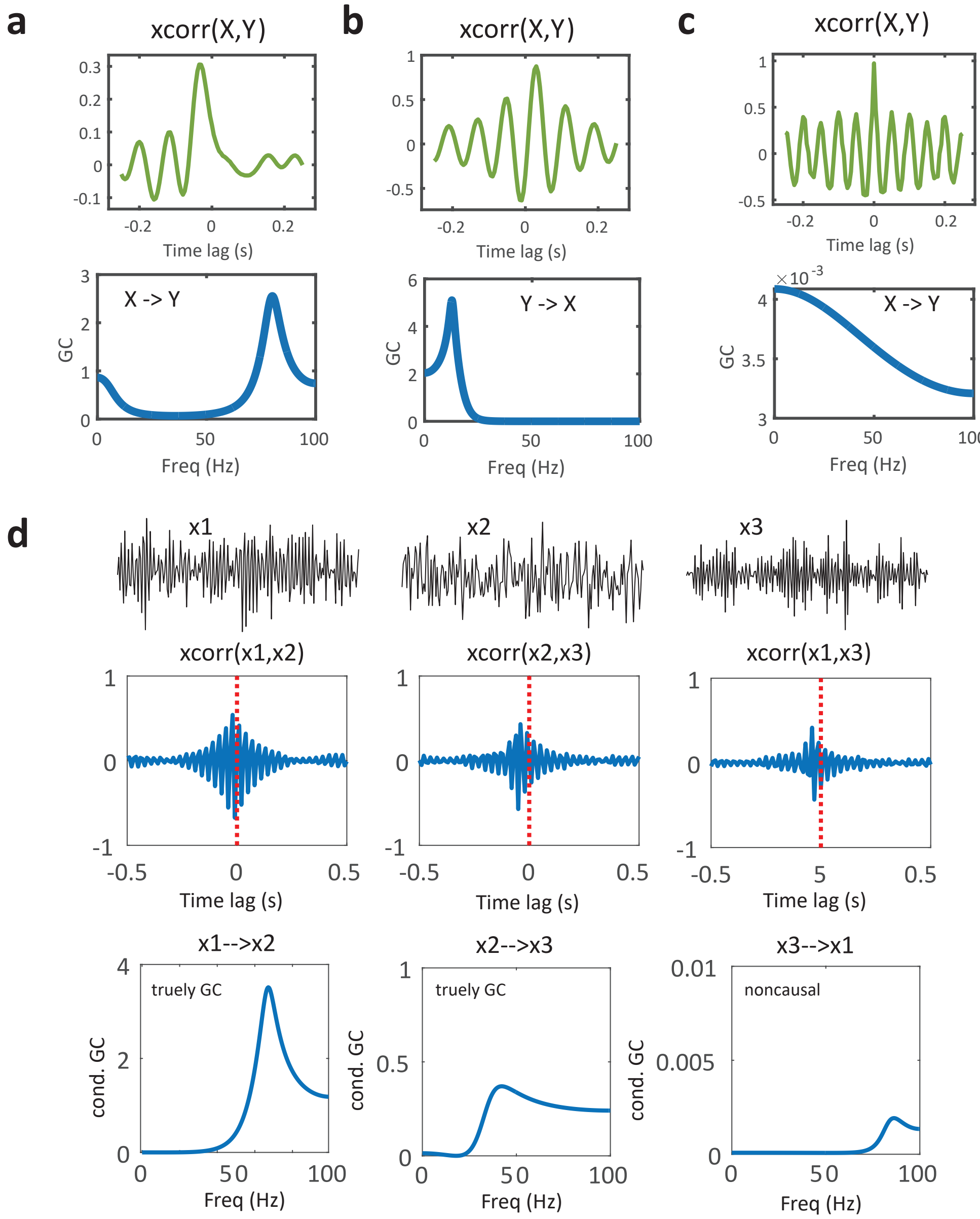

**Figure S1. Lagged cross-correlation (xcorr) between two random variables is neither sufficient nor necessary condition for Granger causality (GC). Here all time-series have the same sampling frequency.**

**(a)** $X$ Granger causes $Y$, with a clear asymmetric xcorr profile.

**(b)** $X$ Granger causes $Y$, but the xcorr profile is nearly symmetric.

**(c)** $X$ and $Y$ has a pronounced xcorr profile, but there is no statistically significant GC.

**(d)** *Top:* Snapshots of time series $\{x_1, x_2, x_3\}$ generated from a VAR(3) based on REF[54]; *Middle:* xcorr profiles; *Bottom:* ground truth conditional SGC ($x_1$ Granger causes $x_2$ and $x_2$ Granger causes $x_3$). In this case, there is no a clear indication of GC from xcorr.

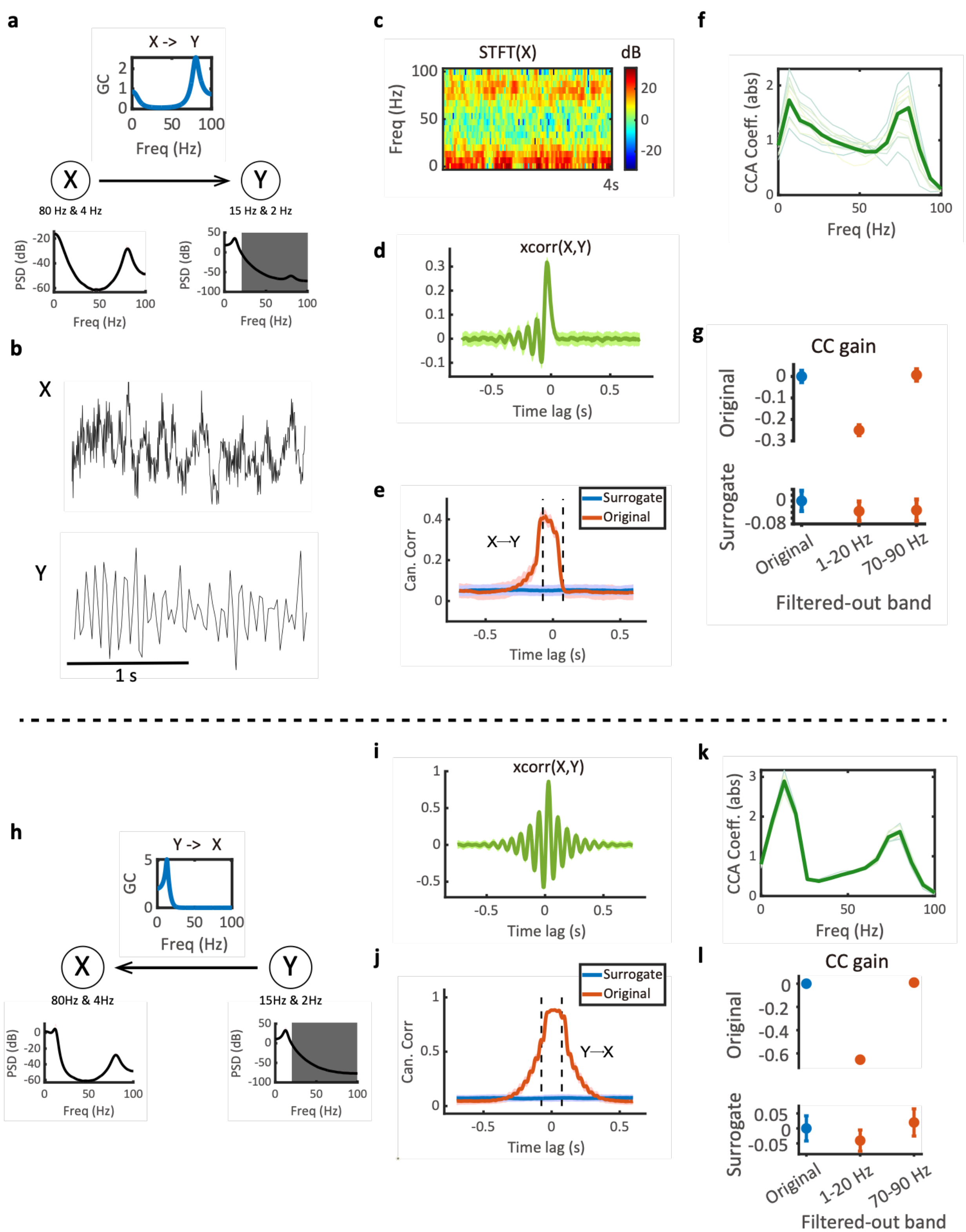

**Figure S2. Computer simulation results on bivariate unidirectional Granger causality (GC).**

**(a)** Illustration of the unidirectional GC ($X \overset{f_x}{\to} Y$) at both low frequency $f_x$ = 4 Hz and high frequency $f_x$ = 80 Hz and the power spectral density (PSD) of two time series. $X$ has a low-frequency component and a high-frequency component, and $Y$ has two resonant frequencies in the low-frequency band. The shaded area in the PSD denotes the filtered-out frequency band above the cut-off frequency (20 Hz) of down-sampled $Y$.

**(b)** The 2-s snapshots of bivariate time series.

**(c)** The 4-s snapshot of short-time Fourier transform (STFT) of $X(t)$.

**(d)** Lagged cross-correlation profile generated from xcorr(X,Y).

**(e)** Lag-CC profile generated from linear MF-TFCCA, which detected directed information flow from $X$ to $Y$. Red trace denotes the estimate derived from the data, and blue trace denotes the estimate derived from the surrogate data. Shade areas denote 95% confidence intervals (n=100). The significant CC profile located beyond the left dashed line suggests causal information from $X$ to $Y$.

**(f)** The MF-TFCCA coefficients computed at a negative lag show a peak in the low-frequency range and another peak in the high-frequency range, corresponding to the two resonant frequencies of $X(t)$: 4 Hz and 80 Hz. The bold red line indicates the trial-averaged coefficients, and the light lines indicate the coefficients estimated from single trials.

**(g)** The CC gain (ΔCC relative to the original unfiltered setting) by band-stop filtering of relevant frequency band in the HF signal $X$. A negative gain suggests greater importance of the putative driving frequency (i.e., 4 Hz) within the filter-out band, whereas a zero gain suggests negligible change in CC of the other putative driving frequency (e.g., 80 Hz).

**(h)** Illustration of the unidirectional GC ($Y \overset{f_y}{\to} X$) at only low frequency $f_y$ = 15 Hz and the PSD of two time series. The shaded area in the PSD denotes the filtered-out frequency band above the cut-off frequency (20 Hz) of down-sampled $Y$.

**(i)** Lagged cross-correlation profile generated from xcorr(X,Y).

**(j)** Lag-CC profile generated from linear MF-TFCCA recovered a directed information flow from $Y$ to $X$.

**(k)** The CCA coefficients (in absolute value) computed from MF-TFCCA at a positive lag showed a higher peak in 15 Hz than in 80 Hz. Despite down-sampling, the driving frequency of $X$ was preserved.

**(l)** The relative CC gain (ΔCC relative to the original unfiltered setting) by band-stop filtering of relevant frequency band in the HF signal $Y$. A negative gain suggests greater importance of the putative driving frequency (i.e., 15 Hz) within the filter-out band, whereas a zero gain suggests negligible change in CC induced by the other putative driving frequency (i.e., 80 Hz).

**a**

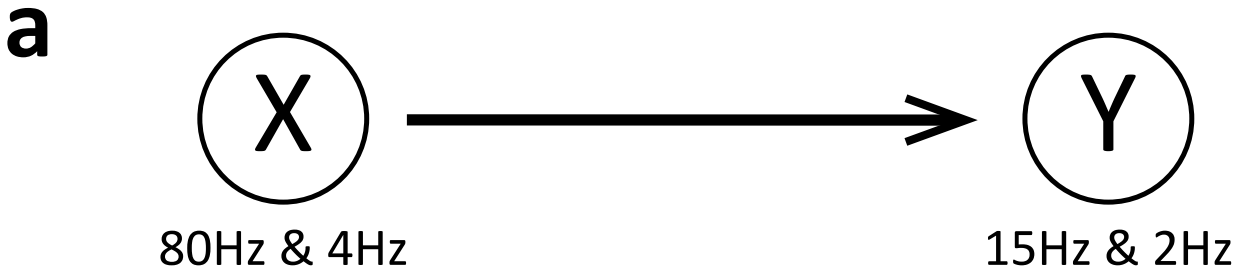

**b**

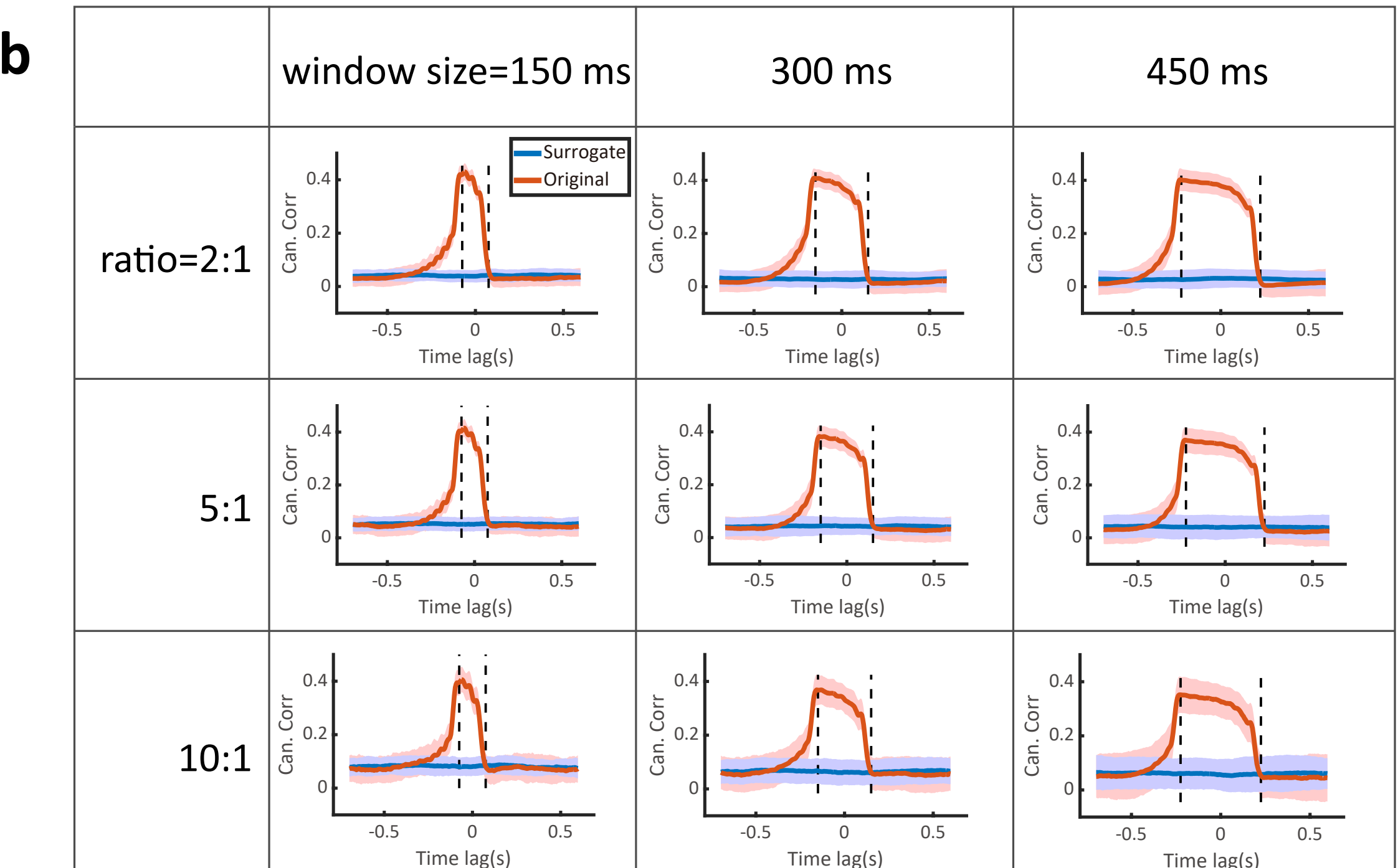

**Figure S3. Comparison of estimated lag-CC profiles from MF-TFCCA with various sampling frequency ratios $\frac{F_x}{F_y}$ and STFT window sizes.**

- **(a)** Illustration of the unidirectional GC system (same as **Fig. S2a**).
- **(b)** The estimated lagged cross-correlation profile under different conditions with various sampling frequency ratios and window sizes in STFT. The significant CC profile located beyond the left dashed line suggests causal information from $X$ to $Y$. Note that the window size determines the detection resolution (i.e., the duration between two vertical dashed lines).

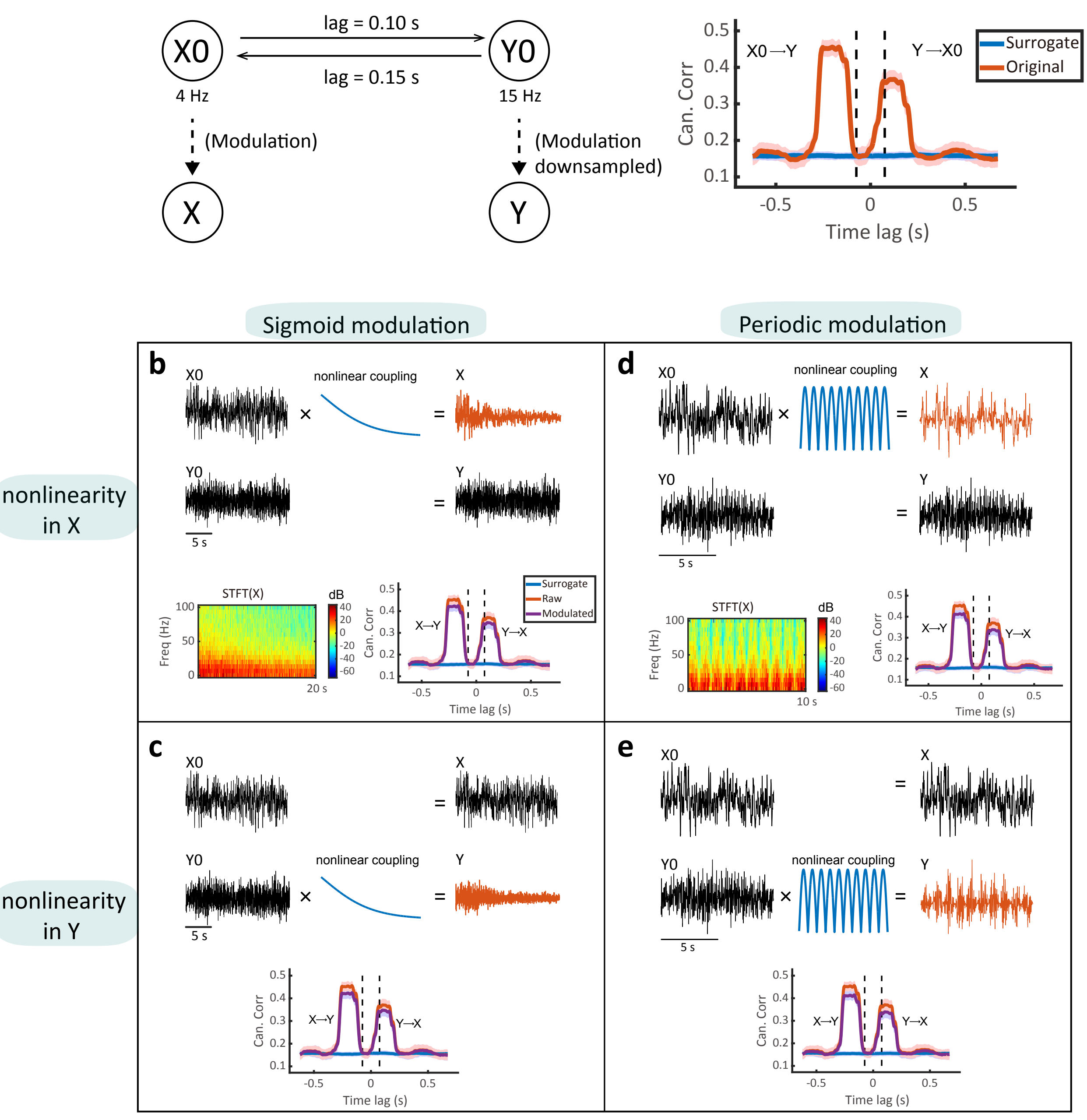

**Figure S4. Computer simulation results on bivariate nonlinear Granger causality (GC) with amplitude modulation.**

**(a)** The original bivariate linear GC system, with $X_0 \overset{f_x}{\rightarrow} Y_0$ at $f_x$ = 4 Hz and $Y_0 \overset{f_y}{\rightarrow} X_0$ at $f_y$ = 15 Hz. The nonlinear causal system variables $X$ and $Y$ are generated by modulating the amplitude $X_0$ or $Y_0$. MF-TFCCA recovered the bidirectional causal structure between $X_0$ and $Y_0$.

**(b)** Nonlinear causal system with sigmoid amplitude modulation. *Top:* the modulated signal $X$ was produced by the raw signal $X_0$ multiplying a sigmoid function, where $Y$ remained intact. *Bottom left:* the STFT spectrum of the modulated $X$ shows the decay in the power of $X$. *Bottom right*: the lag-CC curve revealed bidirectional nonlinear causal relationship between $X$ and $Y$.

**(c)** Similar to panel **b**, but the sigmoid modulation was applied to $Y$. Similar results were found.

**(d)** A nonlinear GC system with periodic amplitude modulation. The modulated signal $X$ was produced by the raw signal $X_0$ multiplying a sine function, where $Y$ remained intact. The STFT spectrum of the modulated $X$ showed a periodic pattern. The lag-CC curve also revealed bidirectional nonlinear causal relationship between $X$ and $Y$.

**(e)** Similar to panel **d**, but the periodic modulation was applied to $Y$. Similar results were found.

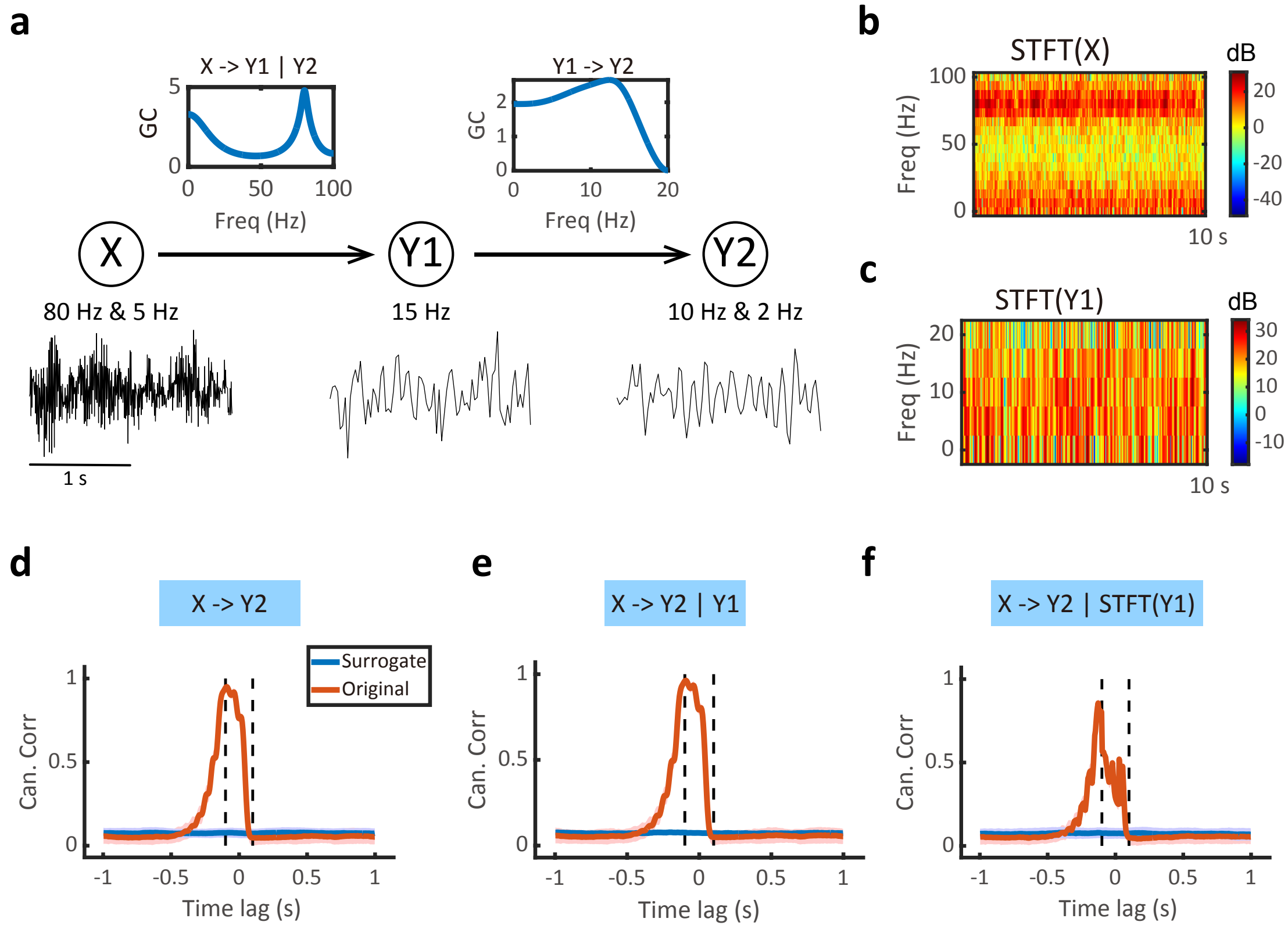

**Figure S5. Computer simulation results on trivariate Granger causality (GC) in a chain system with the intermediate variable in low sampling rate.**

- **(a)** Illustration of the chain system $X \rightarrow Y_1 \rightarrow Y_2$ as well as the conditional GC profiles. Note that $Y_1$ and $Y_2$ have low sampling rates (40 Hz), whereas $X$ has a high sampling rate (200 Hz). The system is similar to the chain system in **Fig. 4** except that $Y_1$ here was down-sampled from $X_2$ in **Fig. 4**.
- **(b)** STFT spectrogram of $X$.
- **(c)** STFT spectrogram of $Y_1$. Note that in order to match the length of STFT($X$), the overlap of STFT window was different, resulting in different temporal resolution.
- **(d)** Lag-CC profile generated from MF-TFCCA without considering $Y_1$, showing directed information flow from $X$ to $Y_2$. Red trace denotes the estimate derived from the original data, and blue trace denotes the estimate derived from the surrogate data. Shade areas denote 95% confidence intervals (n=100 trials).
- **(e)** Similar to panel **d**, but the lag-CC profile was computed based on partializing $Y_1$ in the time domain. The lag-CC magnitude was not reduced by partializing the time series $Y_1$.
- **(f)** Similar to panel **d**, but the lag-CC profile was computed based on partializing the spectrum $\mathrm{STFT}(Y_1)$ in the time-frequency domain. $\mathrm{STFT}(Y_1)$ was computed using the same window size as $\mathrm{STFT}(X)$ and with the same overlap of the moving window as the sampling interval of $Y$. Note that the lag-CC magnitude decreased after partializing $Y_1$ spectrum, suggesting a chain structure in this system. However, the residue was still large due to the information loss of down-sampling $Y_1$.

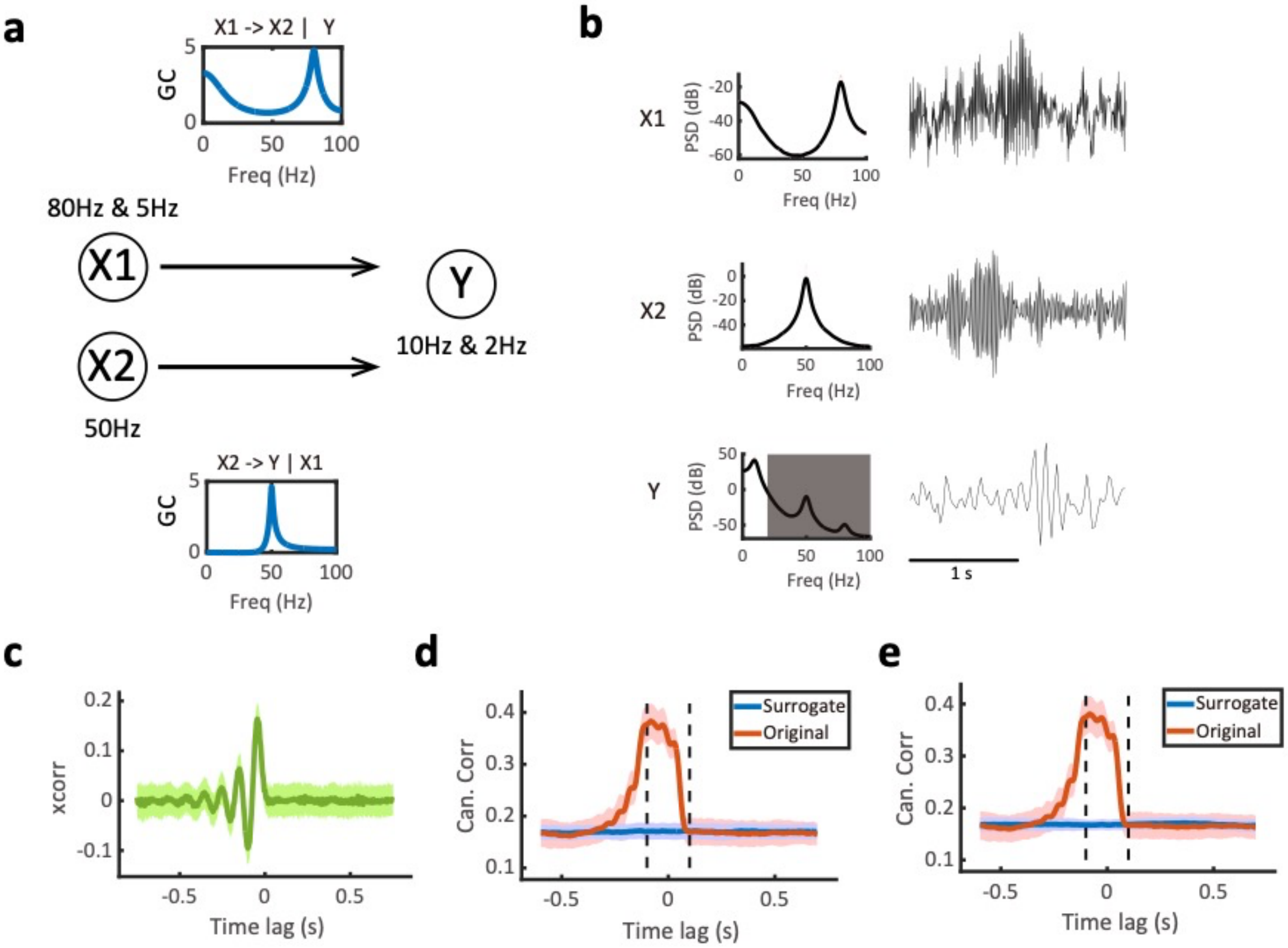

**Figure S6. Computer simulation results on trivariate Granger causality (GC) in a parallel system.**

- **(a)** Illustration of the parallel system $X_1 \rightarrow Y$ and $X_2 \rightarrow Y$ as well as the conditional GC profiles.
- **(b)** 2-s snapshots of trivariate time series and their power spectral density (PSD). Note that $X_1$ and $X_2$ have high sampling rates (200 Hz), whereas $Y$ has a low sampling rate (40 Hz). The shaded area in the PSD denotes the filtered-out frequency band above the cut-off frequency (20 Hz) of down-sampled $Y$.
- **(c)** The lagged cross-correlation profile generated from xcorr($X_1$,Y) without considering $X_2$.
- **(d)** The lag-CC profile generated from MF-TFCCA without partialization of $X_2$, showing a directed information flow from $X_1$ to $Y$. Red trace denotes the estimate derived from the original data, and blue trace denotes the estimate derived from the surrogate data. Shade areas denote 95% confidence intervals (n=100 trials).
- **(e)** The lag-CC profile generated from MF-TFCCA with partialization of $X_2$, showing a similar directed information flow from $X_1$ to $Y$; namely, the driving of $X_2 \rightarrow Y$ did not affect detecting $X_1 \rightarrow Y$.

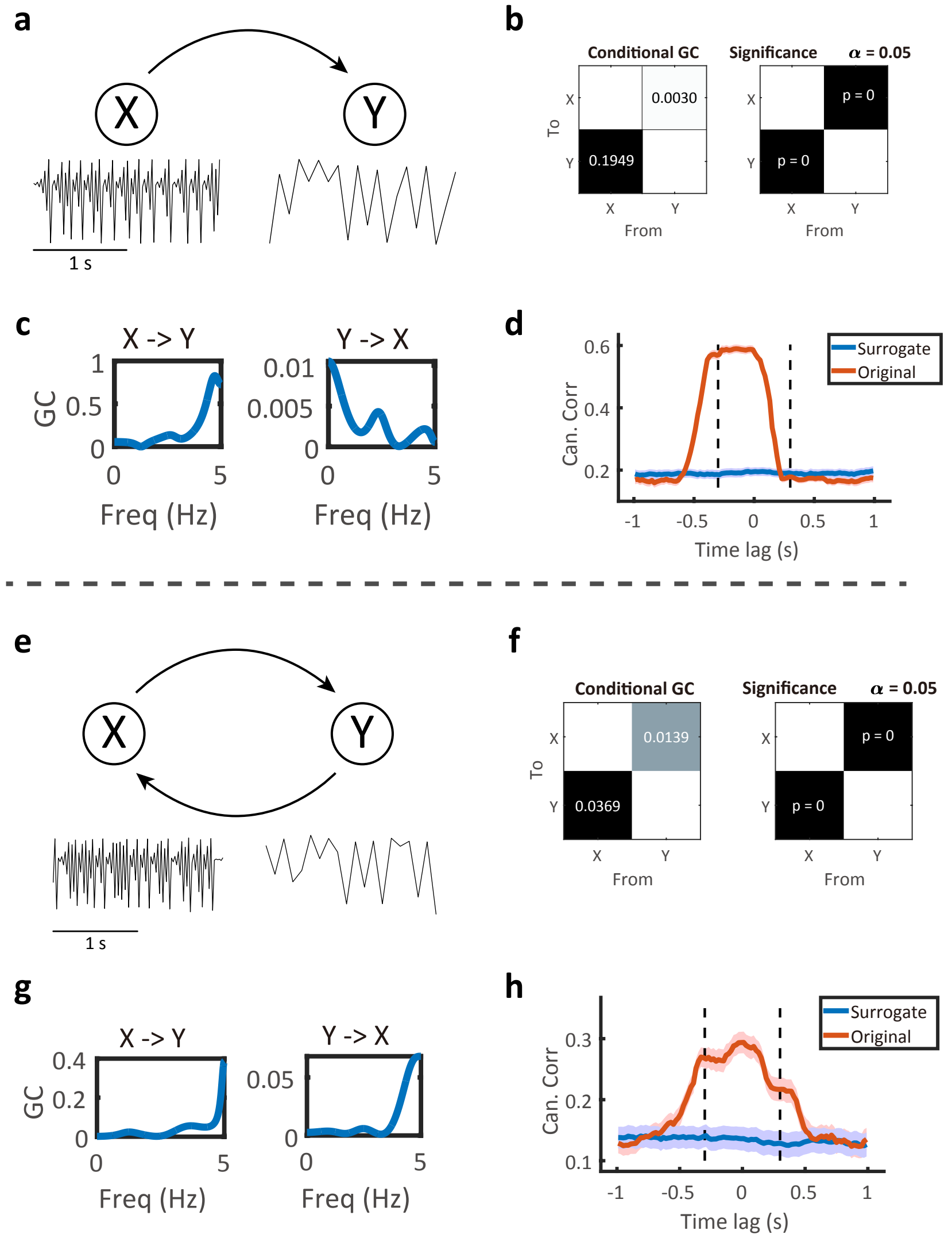

**Figure S7. Computer simulation results on the two-species Logistic model.**

**(a)** Illustration of the unidirectional system $X \to Y$, where $Y$ has a lower sampling rate than $X$ by a factor of 5.

**(b)** Time-domain GC test showing GC value (left) and p-value (right) under a significance level of 0.05.

**(c)** SGC profiles in the frequency domain. A dominant SGC coefficient was found in the $X \to Y$ direction.

**(d)** The lag-CC profile generated from MF-TFCCA discovered a directed information flow from $X$ to $Y$. Red trace denotes the estimate derived from the original data, and blue trace denotes the estimate derived from the surrogate data. Shade areas denote 95% confidence intervals. The significant CC profile located beyond the left dashed line suggests causal information from $X$ to $Y$.

**(e)** Illustration of the bidirectional system $X \leftrightarrow Y$.

**(f)** Time-domain GC test showing GC value (left) and p-value (right) under a significance level of 0.05.

**(g)** Greater SGC coefficients were found in the $X \to Y$ direction than the $Y \to X$ direction.

**(h)** The lag-CC profile generated from MF-TFCCA discovered a bidirectional information flow, as suggested by the significant CC profile located beyond both left and right vertical dashed lines.

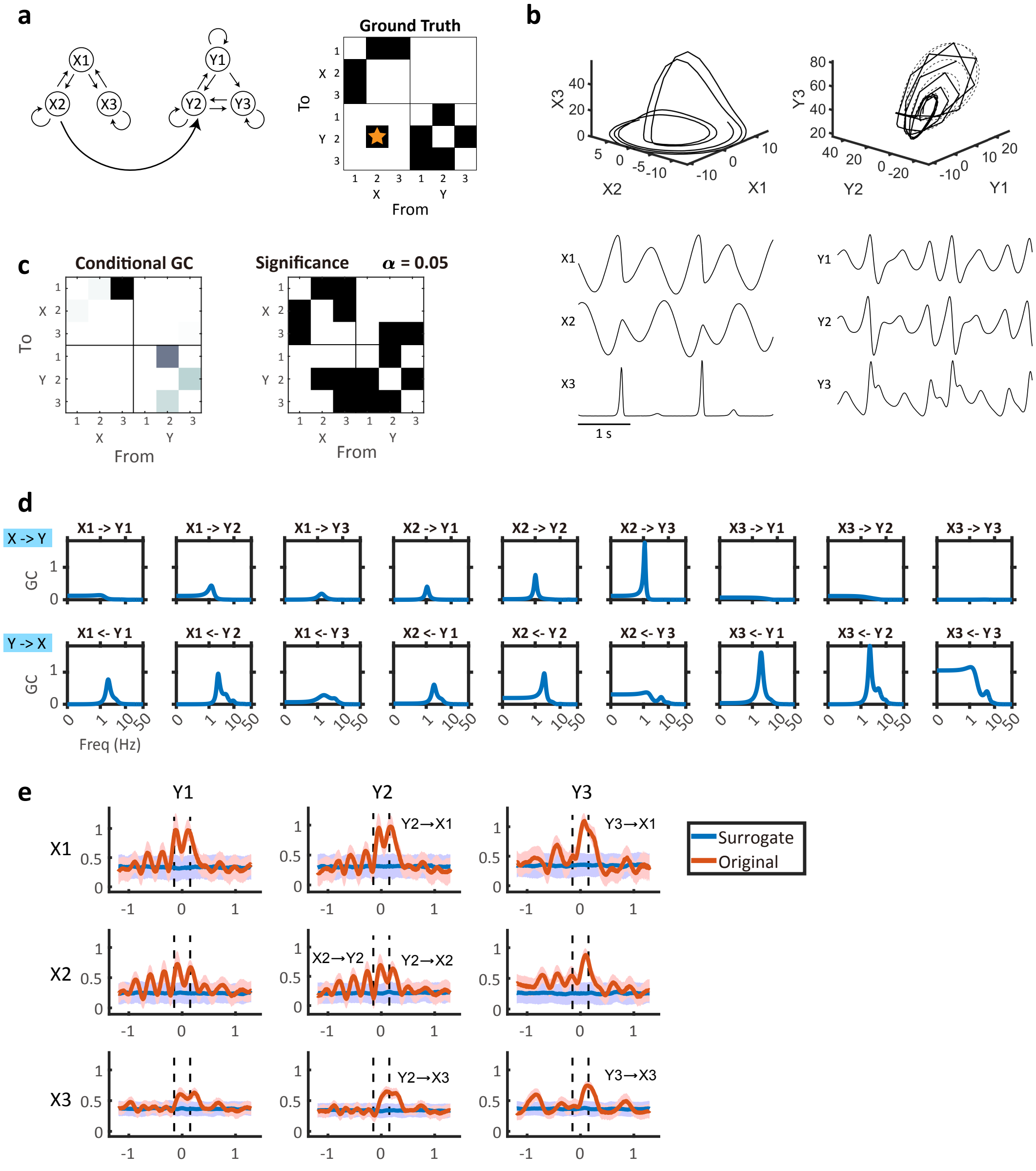

**Figure S8. Computer simulation results on the coupled Rössler-Lorenz system.**

- **(a)** Illustration of the true causal relationship of all 6 observed variables $\boldsymbol{X} = \{x_1, x_2, x_3\}$ and $\boldsymbol{Y} = \{y_1, y_2, y_3\}$. The black entries denote the statistical dependency between variables in terms of system dynamics. The black entry with "★" symbol illustrates the nonlinear coupling $x_2 \rightarrow y_2$ between two systems.
- **(b)** Snapshots of state-space portraits and time series for HF signals $\boldsymbol{X}$ and LF signals $\boldsymbol{Y}$.
- **(c)** Standard GC test in the time domain based on observed time series with equal sampling frequency, which shows the GC values (left) and statistical significance (right).
- **(d)** Estimated SGC in the frequency domain based on observed time series with equal sampling frequency. Note that the frequency axis is shown in the log scale.
- **(e)** The lag-CC profiles for all pairwise MF signals in MF-TFCCA (based on real-valued STFT spectrum).

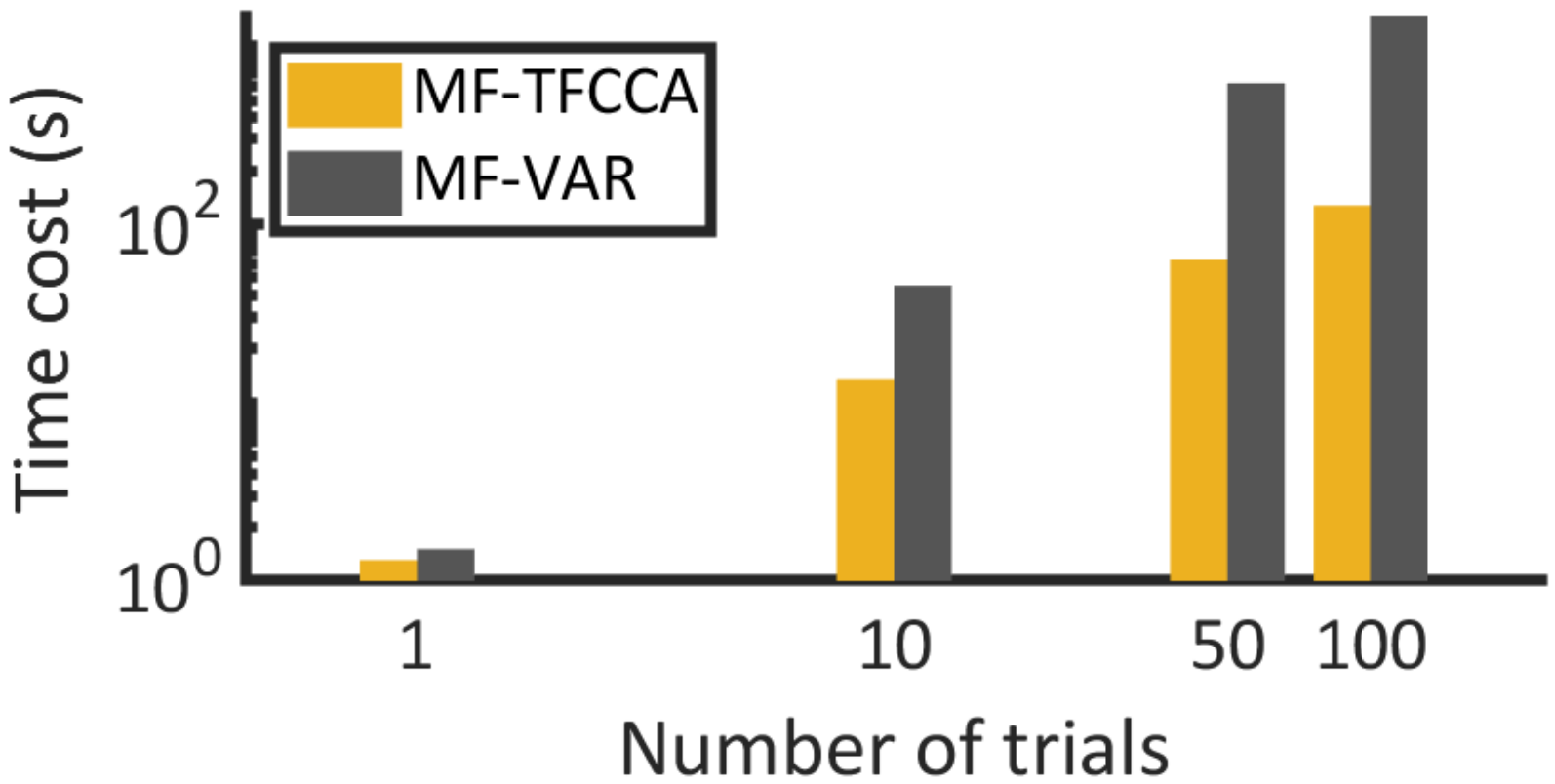

**Figure S9. Benchmark comparison between MF-TFCCA and MF-VAR on computational cost.** In this illustration, we used a simple bivariate VAR(1) system setup with unidirectional GC and a down-sampling factor of 5. Notice that the CPU time cost increased nearly quadratically for MF-VAR in terms of number of trials for a single-lag computation, whereas increased linearly for MF-TFCCA. Note that the plot is in log-log scale. In computer simulations, each trial contained 4000 samples of the HF time series, and 800 samples of the LF time series. The computational cost also included computation of 100 repetitions of surrogate data.

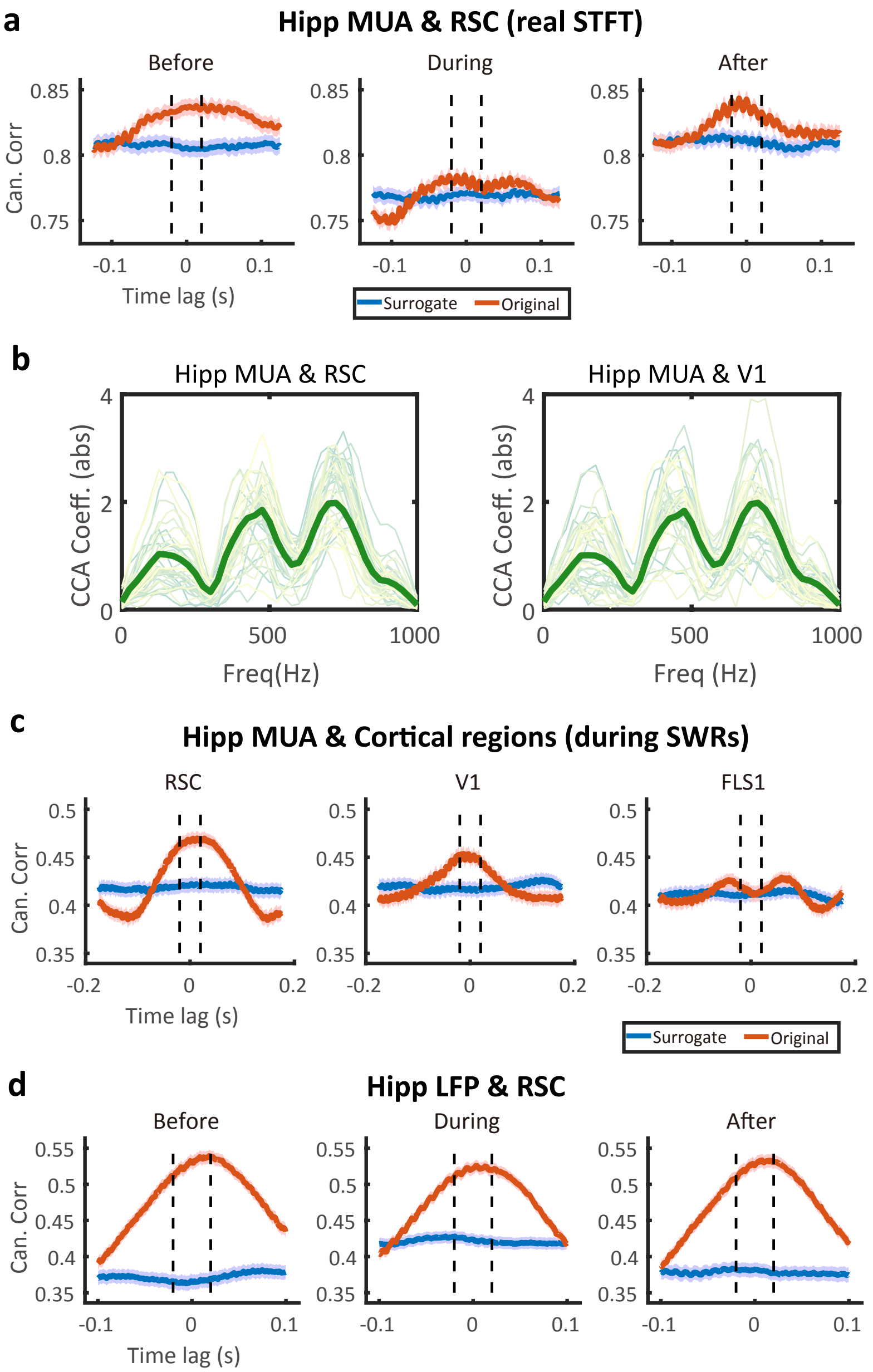

**Figure S10. Inferred directed information flow in mouse hippocampal-neocortical recordings.**

- **(a)** The lag-CC profiles for high-sampled hippocampal MUA and low-sampled RSC activity before, during, and after ripple events. The original CC values (red) were compared against the surrogate data estimates (blue).
- **(b)** CCA coefficients (in absolute value) between hippocampal MUA and RSC activity (left) as well as between hippocampal MUA and V1 activity (right). Thick red lines denote the trial average from single trials (thin colored lines). Multiple peaks were found, revealing the driving frequencies of MUA.
- **(c)** The lag-CC profiles for high-sampled hippocampal MUA and low-sampled RSC, V1, and FLS1 activities during sharp-wave ripples (SWRs).
- **(d)** The lag-CC profiles for high-sampled hippocampal LFP and low-sampled RSC activity before, during, and after ripple events. Similar to the hippocampal MUA result, a bidirectional information flow was identified.